\documentclass[12pt,preprint]{aastex}

\shorttitle{Star formation in AGN host galaxies at $z < 1$}
\shortauthors{C.\ M.\ Pierce, D.\ R.\ Ballantyne and R.\ J.\ Ivison}

\newcommand\q{\par\hangindent 0.5cm\hangafter=1\noindent}

\newcommand\Msun{{\,M_\odot}}

\begin{document}

\title{Radio stacking reveals evidence for star formation in the host galaxies
  of X-ray selected active galactic nuclei at $z<1$}

\author{C.\ M.\ Pierce\altaffilmark{1}, D.\ R.\
  Ballantyne\altaffilmark{1} and R.\ J.\ Ivison\altaffilmark{2,3}}

\altaffiltext{1}{Center for Relativistic Astrophysics, School of
  Physics, Georgia Institute of Technology, 837 State Street, Atlanta,
  GA 30332, USA; christina.pierce@physics.gatech.edu} 
\altaffiltext{2}{UK Astronomy Technology Centre, Royal Observatory,
  Blackford Hill, Edinburgh EH9 3HJ}
\altaffiltext{3}{Institute for Astronomy, University of Edinburgh,
  Blackford Hill, Edinburgh EH9 3HJ}

\begin{abstract}
Nuclear starbursts may contribute to the obscuration of active
galactic nuclei (AGNs). The predicted star formation rates are modest,
and, for the obscured AGNs that form the X-ray background at $z < 1$,
the associated faint radio emission lies just beyond the sensitivity
limits of the deepest surveys. Here, we search for this level of star
formation by studying a sample of 359 X-ray selected AGNs at $z<1$
from the COSMOS field that are not detected by current radio
surveys. The AGNs are separated into bins based on redshift, X-ray
luminosity, obscuration, and mid-infrared characteristics. An estimate
of the AGN contribution to the radio flux density is subtracted from each
radio image, and the images are then stacked to uncover any residual
faint radio flux density. All of the bins containing 24~$\mu$m-detected AGNs
are detected with a signal-to-noise $>3\sigma$ in the stacked radio
images. In contrast, AGNs not detected at 24~$\mu$m are not detected
in the resulting stacked radio images. This result provides strong
evidence that the stacked radio signals are likely associated with
star formation. The estimated star formation rates derived from the
radio stacks range from 3~$\Msun$~yr$^{-1}$ to
29~$\Msun$~yr$^{-1}$. Although it is not possible to associate the
radio emission with a specific region of the host galaxies, these
results are consistent with the predictions of nuclear starburst disks
in AGN host galaxies.
\end{abstract}


\section{Introduction}\label{intro}
Energy released by matter spiraling onto a supermassive black hole (an
active galactic nucleus, or AGN) radiates outward and is often
absorbed by material in the host galaxy. The unified model of AGNs
(Antonucci 1993) suggests a ``torus'' of absorbing material
surrounding the AGN, allowing different levels of observed obscuration
to be potentially explained by the observer's line-of-sight to the
AGN. Models of the torus have ranged from a smooth, donut-like
structure (e.g., Pier \& Krolik 1992) to a clumpy medium that is
filled with clouds of obscuring material, which are randomly
distributed around the AGN (e.g., Krolik \& Begelman 1988; K$\ddot{\rm
  o}$nigl \& Kartje 1994; Elitzur \& Shlosman 2006; Nenkova et
al.\ 2008a,b), but its exact origin and structure are still unknown.

Studies of the hard X-ray background reveal that the Seyfert galaxies
that dominate the background have a Type 2 (obscured) to Type 1
(unobscured) ratio of approximately 4:1 (Gilli et al.\ 2007). This
ratio constrains the structure of the obscuring medium, in that it
requires the absorbing material to cover roughly 80\% of the sky from
the perspective of the black hole. This fact, in turn, implies that
the obscuration must be geometrically thick (e.g., Ibar \& Lira
2007). Explanations for stably inflating such a long-lived structure
include X-ray heating of the outer accretion disk (Chang et
al.\ 2007), absorption by dust of AGN emission and subsequent
re-radiation at infrared (IR) energies (Krolik 2007; Shi \& Krolik
2008), and feedback from supernova and radiation pressure from a disk
of star formation in the nucleus (Fabian et al.\ 1998; Wada \& Norman
2002; Thompson et al.\ 2005; Ballantyne 2008). The latter author found
that a pc-scale starburst with star formation rates (SFRs) $\sim
10$~M$_{\odot}$~yr$^{-1}$ may fuel and obscure a Seyfert galaxy. The
infrared emission from the disk is predicted to be easily detectable
by \textit{Spitzer} and \textit{Herschel}, but this radiation may be
reprocessed in the host galaxy and will be difficult to separate from
AGN heated dust. Alternatively, the radio emission from the nuclear
starburst is not reprocessed by the galaxy and, depending on the SFR
and AGN luminosity, could outshine any non-thermal AGN flux. Assuming
the local radio-IR correlation holds for these redshifts and
star-forming densities, Ballantyne (2008) predicted that the 1.4~GHz
radio flux density from a nuclear starburst was $\sim 10$--$30$~$\mu$Jy at
$z=0.8$. This flux density was just below the sensitivity limits of the
deepest radio surveys available at the time. Interestingly, Bauer et
al.\ (2002) noted that the fraction of radio sources with X-ray
matches increased for the fainter radio objects. More recently,
Ballantyne (2009) found that moderate levels of star formation in
obscured AGNs is required for the AGN population to match the observed
1.4~GHz AGN radio counts. 

Here we perform a simple test of the nuclear starburst disk model for
obscuring the X-ray background AGNs by searching for a clear
association of faint radio flux with X-ray selected AGNs at $z <
1$. The experiment is straightforward: stacking radio images at the
positions of known X-ray selected AGNs that were not individually
detected by the deep radio survey. The AGNs used in the stacks can be
selected based on a range of redshift, X-ray luminosity, or the level
of obscuration. After correcting for the expected AGN radio emission,
the stacks will then reveal whether additional radio emission from
star formation is required for the different AGN samples.  This paper
begins with a description of the multi-wavelength dataset used in this
work and the selection of the sample of radio undetected $z < 1$ AGNs
(Section~\ref{data}). Section~\ref{proc} describes how these AGNs were
split into various samples, the radio stacking procedure, and the
accompanying results. Finally, Section~\ref{conc} provides a short
discussion and our conclusions. Where needed, we use AB magnitudes and
\{$H_0$, $\Omega_{\Lambda}$, $\Omega_{M}$\} $=$ \{$70$, $0.7$,
$0.3$\}.

\section{Data}\label{data}
We use data from the Cosmic Evolution Survey (COSMOS; see the overview
by Scoville et al.\ 2007), an equatorial field encompassing
approximately 2~deg$^{2}$. In particular, we make use of the X-ray and
radio imaging, as well as redshifts determined from optical
spectroscopy and optical photometry. Table~\ref{table:data} provides a
summary of the X-ray and radio observations undertaken as part of
COSMOS. As described below, we select from these datasets the 481 AGNs
at $z < 1$ with estimated unabsorbed $2$--$10$~keV luminosities
$>10^{41.5}$~erg~s$^{-1}$. We then identify the AGNs that are
undetected in the deep VLA 1.4~GHz images of the field. After
excluding two additional AGNs that have potential radio contamination
(Section~\ref{data:radio}), {\it our final sample consists of 359
  radio-undetected AGNs}. To place these results in context, a brief
analysis of the 120 AGNs that are detected in the VLA 1.4~GHz images
is presented in Appendix~A.

\subsection{X-ray observations}\label{data:xray}
Both {\it Chandra} and {\it XMM-Newton} were used to observe the
COSMOS region. Although the {\it Chandra} observations achieved higher
sensitivity (Elvis et al.\ 2009) than the {\it XMM-Newton}
observations, they only covered the central 0.9~deg$^{2}$ of the
field. While the greater sensitivity may be ideal for certain studies,
we choose to take advantage of the larger area (approximately
2.13~deg$^{2}$) observed by {\it XMM-Newton} (Hasinger et al.\ 2007;
Cappelluti et al.\ 2009). These observations reached flux limits of
$1.7 \times 10^{-15}$~erg~s$^{-1}$~cm$^{-2}$ and $9.3 \times
10^{-15}$~erg~s$^{-1}$~cm$^{-2}$ in the soft (0.5--2~keV) and hard
(2--10~keV) X-ray energy bands, respectively (Cappelluti et
al.\ 2009). 

\subsection{Spectroscopic and photometric redshifts}\label{data:redshift}
Sources detected by the {\it XMM-Newton} observations have been
matched to several other surveys of the COSMOS field (Brusa et
al.\ 2010). Most importantly for the current work, this includes
spectroscopic and photometric redshifts, with which we restrict our
sample to X-ray sources at redshifts $0<z<1$. Brusa et al.\ (2010)
provide details of the matching. The spectroscopic redshifts included
in the catalog from Brusa et al.\ (2010) have been gathered from
several sources, including the zCOSMOS
bright and faint catalogs (Lilly et al.\ 2007, 2008, 2009; Lilly et
al., in preparation), the Sloan Digital Sky Survey (SDSS; Kauffmann et
al.\ 2003; Adelman-McCarthy et al.\ 2006; Prescott et al.\ 2006), and
observations taken with the Multi-Mirror Telescope (MMT; Prescott et
al.\ 2006), the DEIMOS instrument on the Keck-II telescope, and the
Magellan/IMACS instrument (Trump et al.\ 2007, 2009). Additionally,
Brusa et al.\ (2010) provide photometric redshifts measured and
originally presented by Salvato et al.\ (2009). Reliable spectroscopic
redshifts are available for 67\% (240/359) of the AGNs in our final
sample. Of these 240 redshifts, 59\% (142/240) come from the zCOSMOS
bright survey, and 26\% (63/240) come from IMACS
observations. Analyses of SDSS, MMT, and Keck/DEIMOS spectra together
provide the remaining 15\% (35/240) of the spectroscopic redshifts. We
use photometric redshifts for the remaining 119 AGNs.

\subsection{Column densities and intrinsic X-ray luminosities}\label{data:nh_lx}
In order to securely identify AGNs based on their X-ray luminosities,
we need to account for the amount of obscuration along our
line-of-sight to the AGN. We estimate this obscuration, quantified by
the neutral hydrogen column densities ($N_{\mathrm{H}}$), using
XSPEC\footnote{http://heasarc.gsfc.nasa.gov/docs/xanadu/xspec/index.html}
version 12.6.0, an X-ray spectral fitting package. Earlier work by
Mainieri et al.\ (2007) also utilized XSPEC to measure column
densities of COSMOS X-ray sources. Even though they followed a
different method than we describe below, a comparison of the 14 X-ray
sources in common between their sample and our AGN sample indicates a
good match. Eleven of the 14 column densities (79\%) match to within
$2\sigma$, where the significance of the match has been estimated
using the uncertainties reported in table 1 of Mainieri et
al.\ (2007).

After calculating the ratio between the hard (2--10~keV) and soft
(0.5--2~keV) observed X-ray fluxes, we use XSPEC to model a redshifted
power-law spectrum with two absorption factors and a photon index of
$\Gamma = 1.9$. Our choice of photon index is based on observations of
unobscured AGNs (e.g., Nandra \& Pounds 1994). The two absorption
factors are the Galactic column towards the COSMOS field ($2.7 \times
10^{20}$~cm$^{-2}$; Mainieri et al.\ 2007) and the absorption in the
observed galaxy. The second factor is placed at the same redshift as
the modeled power-law spectrum. We then increase the value of the
second absorption factor, beginning at $10^{20}$~cm$^{-2}$, until the
ratio between the predicted hard and soft X-ray fluxes matches the
observed ratio to within approximately 1\% of the predicted ratio.

We show the distribution of column densities in
Figure~\ref{fig:nh}. Forty-three sources have been assigned a column
density of $\log(N_{\mathrm{H}}/$cm$^{-2})=20$, indicating minimal
attenuation of the soft-band X-ray flux. Using a Kolmogorov-Smirnov
(K-S) test (Fasano \& Franceschini 1987), we compare the X-ray
luminosity distributions and redshift distributions of the 43 AGNs
having $\log(N_{\mathrm{H}}/$cm$^{-2})=20$ with that of the 316 AGNs
having $\log(N_{\mathrm{H}}/$cm$^{-2})>20$. A K-S test indicates a
scaled maximum deviation ($0<D<1$) between the cumulative
distributions of two data sets (such as the X-ray luminosities of the
two samples). It also provides the significance level $p$ of $D$,
where $D \approx 1$ and $p<<1$ indicates that the two samples differ
significantly.

The X-ray luminosity distribution of the 43 AGNs with column density
$\log(N_{\mathrm{H}}/$cm$^{-2})=20$ is very similar to the
distribution of the remaining AGNs ($D=0.10$, $p=0.83$). The redshift
distributions show a very similar discrepancy ($D=0.10$) and again a
high probability of having been drawn from the same parent population
($p=0.85$). The complete sample has a median column density
$\log(N_{\mathrm{H}}/$cm$^{-2})=22.2$, while the 316 AGNs that have
$\log(N_{\mathrm{H}}/$cm$^{-2})>20$ exhibit a median column density of
$\log(N_{\mathrm{H}}/$cm$^{-2})=22.3$. The distribution that we
observe between obscured AGNs (209/359 $=$ 58\%;
$\log(N_{\mathrm{H}}/$cm$^{-2})>22$) and unobscured AGNs (150/359 $=$
42\%; $20\le\log(N_{\mathrm{H}}/$cm$^{-2})\le 22$) is consistent with
established observations of the ratio between obscured and unobscured
AGNs (e.g.\ Maiolino \& Rieke 1995; Ho et al.\ 1997), indicating that
our use of a constant power-law slope of $\Gamma=1.9$ does not
adversely affect our conclusions. Of course, this sample is missing
Compton thick AGNs, those that have column densities
$\log(N_{\mathrm{H}}/$cm$^{-2})>24$. Therefore, our results only apply
to the Compton thin population of AGNs; however, Compton thick objects
seem to be a small fraction of the AGN population (Malizia et
al.\ 2009), and, at these redshifts, may only be associated with major
mergers and interactions (Draper \& Ballantyne 2010; Treister et
al.\ 2010), and the nuclear starburst disk model would not be
applicable.

XSPEC is then used once again to estimate the {\it unabsorbed}
2--10~keV X-ray fluxes, using the same spectral model and parameters
and setting the source column densities to zero. We then derive the
necessary k-correction for the X-ray fluxes by integrating over a
power-law spectrum, which results in the following k-correction:
\begin{equation}
f_{\rm rest} = f_{\rm obs}(1 + z)^{-(\alpha_{x} - 1)},
\end{equation}
where $\alpha_{x} = \Gamma - 1$, and we again use $\Gamma =
1.9$. Using the k-corrected, unabsorbed flux and the luminosity
distance of each individual source, we finally calculate the
intrinsic, rest-frame 2--10~keV luminosities. All of the unabsorbed
fluxes are less than a factor of two greater than the corresponding
absorbed fluxes, with a median correction factor of 1.01. Therefore,
the correction for absorption does not lead to a large difference
between the observed and the intrinsic X-ray luminosities. Panel (a)
of Figure~\ref{fig:lx_z} shows the intrinsic 2--10~keV X-ray
luminosities of our AGN sample, as a function of redshift. The black
curve indicates the hard-band flux sensitivity limit for the survey
(see table 2 from Cappelluti et al.\ 2009), and the gray arrows
indicate the 108 X-ray sources (30\% of our AGN sample) that do not
have significant hard-band detections. The X-ray luminosities for
these sources are derived from absorption-corrected upper limits to
the hard-band X-ray fluxes (Brusa et al.\ 2010). The redshift
distribution, peaking at $z \sim 0.35$ and $z \sim 0.85$, is shown in
panel (b) of Figure~\ref{fig:lx_z}, and is very similar to the one
found by Strazzullo et al.\ (2010) for faint radio sources with flux densities
between 16 and 30~$\mu$Jy. Thus, these X-ray selected AGN seem to be
distributed in redshift in a similar manner to faint radio sources.

\subsection{Radio observations}\label{data:radio}
Radio observations of the full 2 deg$^{2}$ COSMOS field, at 1.4~GHz
(21~cm), comprise the VLA-COSMOS Large Project (Schinnerer et
al.\ 2004, 2007; Bondi et al.\ 2008). Approximately 3600 sources have
been detected at a minimum 4.5 $\sigma$ detection significance, and
the sensitivity of the observations reaches $f \approx$
11~$\mu$Jy/beam. 

We search for radio counterparts to our initial sample of 481 $z < 1$
X-ray selected AGNs using the coordinates provided in the X-ray and
radio catalogs. Searching within a circle of radius 2\arcsec\ centered
on each X-ray source, we find radio counterparts to 120 of the AGNs
with redshifts $z<1$ (25\%; 120/481). The most distant radio
counterpart identified using this search radius is centered
1.38\arcsec\ from the center of the X-ray source. Park et al.\ (2008)
used a less conservative search radius of 2.5\arcsec\ to match X-ray
sources to IR sources. If instead we use the search radius used by
Park et al., one of the X-ray sources has two potential radio matches,
at distances of 2.48\arcsec\ and 0.06\arcsec. In such a case, we would
select the nearer source as the match. A brief analysis of the
properties of these radio-detected AGNs is presented in Appendix~A. 

The radio images and fluxes of all remaining 361 AGNs are then
visually examined. We do not find any that are located near known
radio sources. The radio flux density measurements identify an additional two
X-ray selected AGNs as possible radio sources. A visual inspection of
the radio images suggests that both AGNs may be associated with true
radio sources and are excluded from our sample. Thus, we arrive at 359
radio-undetected X-ray selected AGNs at $z < 1$.

\subsection{Mid-infrared observations}\label{data:ir}
Although not used for source selection, it is interesting to consider
the mid-IR properties of our AGN sample. Sanders et al.\ (2007)
describe S-COSMOS, the mid- to far-IR survey of the COSMOS field,
undertaken with the {\it Spitzer Space Telescope} ({\it Spitzer}). The
entire field was observed at wavelengths from 3.6~$\mu$m to
160~$\mu$m. Brusa et al.\ (2010) identified the 24-$\mu$m counterparts
of the COSMOS X-ray sources, using a mid-IR catalog presented by Le
Floc'h et al.\ (2009). Eighty-four percent (301/359) of our AGNs have
identified 24-$\mu$m counterparts, indicating that star formation is
likely occurring in these galaxies.
In principle, the 24~$\mu$m flux densities could be used to directly estimate
the SFRs of these galaxies (e.g., Rieke et al.\ 2009). However, the
presence of a luminous AGN will also cause mid-IR emission due to dust
exposed to the AGN. The AGN contribution to the 24~$\mu$m flux density
will be
difficult to estimate as it strongly depends on the geometrical and
physical properties of the dust in the nuclear environment (e.g.,
Nenkova et al.\ 2008; Hatziminaoglou et al.\ 2009).

The column density distributions of the 24~$\mu$m-detected and
undetected AGNs are shown in Figure~\ref{fig:nh_ir}, where the solid
gray histogram represents the 24~$\mu$m-undetected systems (58 AGNs)
and the open histogram represents the 301 AGNs that are detected at
24~$\mu$m. We find a low probability ($p=0.01$) that these are drawn
from the same population and a maximum deviation between the
distributions of $D=0.23$. X-ray luminosities
(Figure~\ref{fig:lx_z_ir} (a)) show a similar deviation ($D=0.20$)
between the 24~$\mu$m-detected and undetected samples and a slightly
larger probability of having been drawn from the same population
($p=0.04$). Figure~\ref{fig:lx_z_ir} (a) reveals a relative lack of
24~$\mu$m-undetected AGNs at the lowest redshifts of our sample ($z <
0.25$). Statistically, however, the redshifts shown in panel (b) of
Figure~\ref{fig:lx_z_ir} are very similar, with a 97\% probability of
coming from the same parent population; the maximum deviation is
correspondingly low ($D=0.07$). In summary, we find that the
24~$\mu$m-detected and undetected samples represent different
sub-populations within the AGN population. As they seem to be a
different population, the 24~$\mu$m-undetected AGNs are treated
separately in the radio-stacking exercise.

\section{Analysis and Results}\label{proc}

\subsection{AGN sub-samples}\label{proc:samples}
We split our AGN sample into several sub-samples based on various
combinations of the following: redshift, detection or non-detection of
24~$\mu$m flux, neutral hydrogen column density, and X-ray
luminosity. All samples are first separated by detection or
non-detection of 24~$\mu$m flux (see Section~\ref{data:ir}) and then
by nuclear obscuration, using a column density
$\log(N_{\mathrm{H}}/$cm$^{-2})=22$, where, following the typical
convention, objects are considered obscured if they are observed
through a $\log(N_{\mathrm{H}}/$cm$^{-2}) > 22$. The samples are then
split by redshift and/or X-ray luminosity. After trying a variety of
sample sizes and grouping strategies, we conclude that samples
containing 55-70 galaxies are best suited for the current work and
that comparing samples of similar size greatly facilitates
interpretation of our results.  

Table~\ref{table:samples} lists the criteria used to create our
sub-samples. The first column provides the ID of each sample, matching
references to specific samples throughout the rest of this paper. The
second column indicates the number of galaxies in each sample. Columns
(3)-(5) list the minimum, maximum, and median redshifts of the
samples, and columns (6)-(8) list the minimum, maximum, and median
X-ray luminosities. We only experiment with a few samples of the
24~$\mu$m-undetected AGN host galaxies, because, as seen below, the
stacked radio images do not result in a positive detection.  

\subsection{AGN contribution to the radio flux density}\label{proc:residuals}
Even though our sample consists of radio undetected AGNs, that does
not imply a complete lack of radio emission from the accreting black
hole. As we are searching for radio emission from star formation in
these AGNs, we need to correct for the AGN contribution to whatever
faint radio flux density exists. Based entirely on observations of local radio
quiet AGNs (Terashima \& Wilson 2003), Ballantyne (2009) used the
following relationship to estimate the rest-frame 5 GHz luminosity
from the known X-ray luminosity:
\begin{equation}
\label{eq:rx}
R_{X} = \left\{\begin{array}{ll}
-0.67 \log L_{X} + 23.67 & 41.5 \le \log L_{X} \le 43 \\
-5                       & 43 < \log L_{X} \le 44 \\
\log L_{X} - 49          & 44 < \log L_{X} \le 45 \\
-4                       & \log L_{X} > 45,
\end{array}
\right.
\end{equation}
where $R_{X} = \log[\nu L_{\nu}({\rm 5 \ GHz})/L_{X}]$ (note that
there is a typographical error in the Ballantyne [2009] definition of
$R_{X}$). The 1.4~GHz luminosity is then calculated assuming a
spectral index of $\alpha=0.7$, where $L_{\nu} \propto
\nu^{-\alpha}$. Finally, the conversion from luminosity to flux density is 
\begin{equation}
\label{eq:flux}
S_{\nu} = \frac{L_{\nu} (1+z)^{1-\alpha}}{4\pi D_{L}^{2}},
\end{equation}
where $S_{\nu}$ is the observed 1.4 GHz flux density in units of
W~m$^{-2}$~Hz$^{-1}$, $L_{\nu}$ is the rest-frame 1.4 GHz luminosity
in units of W~Hz$^{-1}$, and $D_{L}$ is the luminosity distance in
meters. For our sample, the median redshift is $z=0.740$, and the
median predicted 1.4 GHz flux density is $f=6.57 \ \mu$Jy. As there is a wide
dispersion in the radio spectra of AGNs, we checked how our results
depend on the value of $\alpha$. A spectral index $\alpha=0$ decreases
the median predicted flux density to $f=3.8~\mu$Jy, and a steeper spectrum
($\alpha=1$) increases the predicted flux density to $f=8.3~\mu$Jy. All three
values are well below the flux density limit of the COSMOS radio survey (see
Table~1)\footnote{For comparison, the correlation between radio
    luminosities and X-ray luminosities described by Panessa et
    al.\ (2007) for a sample of nearby low luminosity Seyfert galaxies
    predicts a median 1.4 GHz flux density of $58~\mu$Jy. This large value is
    due to the fact that low luminosity AGNs tend to be more luminous
    at radio frequencies (e.g., Ho \& Ulvestad 2001; Nagar et
    al.\ 2002) than higher luminosity Seyferts (as parameterized in
    Eq.~\ref{eq:rx}). Thus, the correlation based on work by Terashima
    \& Wilson (2003) best represents the AGN contribution to the radio
    emission for our sample of AGNs.}.

To remove the estimated AGN contribution to the radio flux density for each
source, we first create a radio PSF by stacking a sample of COSMOS
galaxies from the COSMOS photometry catalog (Capak et al.\ 2007). We
select galaxies that have photometric redshifts $0<z<1$ (to match our
AGN sample), $K$ band magnitudes $20<K<22$, and uncertainties on the
$K$ band measurements $K_{\rm err}<0.5$. Galaxies outside the
radio-imaged region are excluded, as are any sources flagged as having
questionable photometry. We then stack the 23,941 galaxies that meet
these criteria, and the resulting image is our radio PSF. 

For each AGN in our sample, we scale the PSF so that a measurement of
its integrated flux density matches the integrated flux density estimated from the
X-ray luminosity of the AGN. We then subtract the scaled PSF image
from its corresponding AGN radio image, and the residual image
represents the contribution to the observed radio emission from star
formation alone. It is these residual images that are then stacked
using the samples listed in Table~\ref{table:samples}. If the AGN
dominates any faint observed radio emission, then the stack will
reveal no additional radio flux associated with star
formation. However, if one of the AGN sub-populations is experiencing
star formation, the integrated flux and detection significance for the
residual stack may reveal this.

\subsection{Stacking procedure}\label{proc:aips}
Two methods are commonly used to average together the input images to
a stack, and some authors use both methods and then compare the
results. The first of the two methods is a mean or a noise-weighted
mean of the images (e.g., de Vries et al.\ 2007; Ivison et al. 2007;
White et al.\ 2007). This method is preferred for a population that
exhibits a flux distribution (radio flux density, in this case) that is
approximately Gaussian. For samples that are likely to have
non-Gaussian flux distributions, such as our sample of AGNs, a median
or a clipped mean is often used (Wals et al.\ 2005; Boyle et
al.\ 2007; de Vries et al.\ 2007; White et al.\ 2007; Carilli et
al.\ 2008; Hodge et al.\ 2008, Dunne et al.\ 2009). Karim et
al.\ (2011) present a detailed explanation of the statistics
associated with both stacking methods. Because there is no obvious
flux below which the remaining fluxes in our samples follow a Gaussian
distribution, we combine the individual residual radio images for each
sample using a median stacking method.

Each of the stacked images are then analyzed using the AIPS task
\verb|JMFIT| and fluxes are measured with a single two-dimensional
elliptical Gaussian that is fit to a 12 pixel x 12 pixel region
(4.2\arcsec\ x 4.2\arcsec). This analysis region is 
centered at the image center, which corresponds to the locations of
the X-ray sources.
 In order to estimate measurement uncertainties, the AIPS task
 \verb|JMFIT| calculates the root-mean-square of each image within a
 specified radius of the center of the analysis region; we use a
 15-pixel radius. The task runs for up to 1000 iterations, stopping as
 soon as it converges on a solution. The output includes the peak and
 integrated radio flux densities, along with estimated uncertainties on each
 measurement. We define the signal-to-noise ratio (S/N) of each
 measurement as the peak flux density divided by its corresponding
 uncertainty.

\subsection{Stacking results}
\label{proc:stacks}
Figure~\ref{fig:stamps} shows the 15 stacked residual radio images, as
listed in Table~\ref{table:samples}; the PSF is also shown. We
indicate on the figure the 12 pixel x 12 pixel regions that are used
by the AIPS task \verb|JMFIT| to measure the peak and integrated flux
densities of the stacked images. The first 12 samples feature a
visually identifiable radio source, while the last three images
represent the three samples that are not detected at 24~$\mu$m and
show an apparent lack of radio flux. The colors of the latter three
images are inverted to more clearly show the effect of subtracting the
PSF.

Table~\ref{table:results} provides flux densities, detection significances,
and SFRs for the residual radio image stacks. The first column lists
the sample ID, matching those provided in
Table~\ref{table:samples}. Column (2) lists the integrated flux densities of
the stacked samples, together with the associated uncertainties. In
column (3), we list the S/N of each stacked sub-sample. Star formation
rates calculated following equation (6) of Bell (2003) are provided in
column (4) for the samples having non-negative integrated flux densities. The
SFR uncertainties are based on the uncertainties measured with the
integrated flux densities.

As already suggested by Figure~\ref{fig:stamps}, the first 12 samples
show evidence of radio emission, with the S/N ranging from 3.1$\sigma$
to 7.6$\sigma$. Note that the two samples with the highest S/N -- 2z
and 4z -- contain the most AGNs; the remainder of the ``2'' and ``4''
series are sub-sets of the 2z and 4z samples, respectively. When
making catalogs of radio sources, detections with a high significance
are required to ensure the exclusion of false detections (e.g.,
Richards 2000; Biggs \& Ivison 2006; Ivison et al.\ 2007; Schinnerer
et al.\ 2007; Morrison et al.\ 2010). However, when using the stacking
method, we know the locations of the potential sources, which allows
for less strict criteria for a positive detection. All of our
24~$\mu$m-detected samples are detected with S/N $>3$, indicating a
significant radio detection in the stacked image.

In contrast, the 24~$\mu$m-undetected samples show no positive radio
emission, independent of sample size. The integrated flux densities are
negative, but $\lesssim 2\sigma$ from zero. These negative flux densities
might be evidence that the AGN correction that is subtracted from each
image is too large. To test this possibility these three radio stacks
were recalculated with no AGN correction. The resulting stacked images
have a negative integrated flux density for the unobscured sample (``3''), and
positive flux densities for the complete (``1+3''; $2.3\sigma$) and obscured
(``1''; $3.2\sigma$) samples. The unobscured sub-sample (``3'') only
has 14 objects and therefore the null result for this group of AGNs is
likely directly related to the relatively small number of AGNs that
comprise the stack. The complete sample and the obscured sub-sample do
result in weak positive radio detections, yet these disappear if the
AGN correction is performed prior to stacking (Table~3). The estimated
AGN radio flux density depends only on X-ray luminosity (Eq.~\ref{eq:rx}), so
the mean correction for the obscured 24~$\mu$m-undetected sample with
$\langle \log L_X \rangle =43.33$ will be very similar to the one for
the ``2ze'' sample (with $\langle \log L_X \rangle =43.44$), and the
mean correction for the complete sample (with $\langle \log L_X
\rangle =43.19$) will be similar to the one for the ``2zd'' sample
(with $\langle \log L_X \rangle =42.99$). Both the ``2ze'' and the
``2zd'' samples have significant positive detections in the residual
stacks with similar numbers of objects as the 24~$\mu$m-undetected
stacks. If the predicted AGN radio flux density was overestimated than it
would have a comparable effect in all of these samples. Therefore, we
can conclude that the objects in the 24~$\mu$m-detected samples must,
on average, be stronger radio emitters than the 24~$\mu$m-undetected
AGNs, with the additional flux arising from star formation. However,
as described in Sect.~\ref{data:ir}, the 24~$\mu$m-undetected AGNs
seem to arise from a different population than the ones that are
detected at 24~$\mu$m. It is thus possible that the negative flux densities
found in the stacked images of the 24~$\mu$m-undetected AGNs may
indicate that this population of AGNs produces weaker nuclear radio
fluxes than the 24~$\mu$m-detected AGNs, and our AGN correction
(Eq.~\ref{eq:rx}) is an overestimate for this population.

In Figure~\ref{fig:sfr}, we show the SFRs of our samples as a function
of redshift (upper panel) and X-ray luminosity (lower panel). The full
obscured and unobscured samples are labeled with ``2z'' and ``4z'',
respectively. In the upper panel the other symbols represent the
samples that depend on redshift and are labeled with their median
X-ray luminosities, while the lower panel features the samples
dependent on luminosity and are labeled with their median
redshifts. The estimated SFRs are in the range $3.3 \leq {\rm
  SFR}/M_{\odot}$~yr$^{-1} \leq 29$, which are exactly the rates
predicted by nuclear starburst disk models (Ballantyne 2008).
Unsurprisingly, we observe an increase in the SFR with redshift,
consistent with the peak of star formation and supermassive black hole
growth observed at $z \sim 1$ (e.g., Le Floc'h et al.\ 2005). As there
is a correlation between redshift and luminosity
(Figure~\ref{fig:lx_z}), this correlation also appears in the
luminosity panel. Comparing the SFRs of the three higher redshift
samples that have similar median luminosities shows that both the
obscured and the unobscured AGNs have similar levels of star
formation.

Our choice of a spectral index $\alpha=0.7$ for the AGN radio emission
also affects the SFR estimates. We recalculate the SFRs for the 12
samples that have 24~$\mu$m detections using spectral indices of
$\alpha=0$ and $\alpha=1$. In both cases, the resulting star formation
rates are typically lower than those determined for $\alpha=0.7$, but
the deviations from the SFRs reported in Table~\ref{table:results} are
less than the uncertainties derived from the integrated radio flux density
measurements.

\section{Discussion and Conclusions}\label{conc}
The AGN population at $z < 1$ is predominately comprised of obscured
objects with Seyfert luminosities (e.g., Ueda et al.\ 2003), yet the
evolutionary state of these galaxies is largely unknown. At these
luminosities, it is possible that a significant number of AGNs are not
fueled by the violent mergers of massive galaxies, but are undergoing
a more leisurely mode of galaxy assembly (e.g., Ballantyne et
al.\ 2006; Hasinger 2008; Lutz et al.\ 2010). In this case, gas will
be slowly fed into the nuclear regions where significant star
formation will occur en route to the black hole accretion disk. It is
possible that such a nuclear starburst could provide a location for
the X-ray obscuration seen in most of these AGNs (Ballantyne 2008). In
this paper, we searched for evidence for this star formation process
by considering the faint radio emission from 481 $z < 1$ X-ray
selected AGNs in the COSMOS field. The X-ray properties (i.e.,
obscuration and luminosity) of this sample were consistent with the
AGNs that dominate the X-ray background and can be fueled and obscured
by nuclear starbursts.  

As expected from the nuclear starburst model, well over half of these
AGNs (359) were undetected in the VLA 1.4~GHz COSMOS survey, so radio
stacking techniques were employed to search for faint radio emission
in samples that were selected based on redshift, X-ray luminosity, and
obscuration. Objects with 24~$\mu$m detections (301 out of the 359)
were considered separately from those which were undetected at
24~$\mu$m. An estimate for the nuclear AGN radio flux density was subtracted
from each image prior to stacking. For AGNs with a 24~$\mu$m
detection, the stacked images had a positive radio source
corresponding to SFRs of $3.3 \leq {\rm SFR}/M_{\odot}$~yr$^{-1} \leq
29$, consistent with the predictions of a nuclear starburst
(Ballantyne 2008). As expected, the SFRs increased with the redshift
of the samples, but there was no clear dependence on either luminosity
or X-ray obscuration, consistent with the unified model of AGNs and
the starburst disk model. 

The radio stacking was pursued because it was predicted to be a
potentially cleaner test for the presence of star formation than the
mid-infrared. It is interesting, therefore, to compare the SFRs
obtained from the stacking exercise to the ones calculated directly
from the 24~$\mu$m flux densities. Brand et al.\ (2006) estimates that objects
with $f_{24\ \mu\mathrm{m}} < 1.2$~mJy are dominated by
starbursts. Using this criterion, 95\% (287/301) of the 24~$\mu$m
sources matched to our AGNs appear to be dominated by star formation
rather than the AGN, although the AGN will still contribute an unknown
amount of flux to many of these 24~$\mu$m
sources. Figure~\ref{fig:ir_flux} plots the distribution of SFRs
calculated directly from the 24~$\mu$m flux density (Reike et al.\ 2009) for
all 301 AGN host galaxies. The median SFR is 29~$\Msun$~yr$^{-1}$
which is the same as the maximum SFR found in the radio stacks. This
result implies that the AGN contributes a non-negligible fraction of
the 24~$\mu$m flux in at least 50\% of these galaxies. 

The SFRs derived from the radio stacks are consistent with
  measurements of both active and inactive host galaxies at these redshifts. Mullaney et al.\ (2011) recently studied a sample of X-ray selected
AGNs at redshifts $z<3$, combining deep, high-resolution {\it
  Herschel} observations at 100~$\mu$m and 160~$\mu$m with spectral
energy distribution templates (Chary \& Elbaz 2001) to estimate the
total IR luminosities and the SFRs. For the subset of their AGNs at
the redshifts and X-ray luminosities of the current study, Mullaney et
al.\ (2011) found SFRs that are very similar to the SFRs reported
herein. Similarly, Noeske et al.\ (2007) measured SFRs with
optical/UV/IR techniques of massive inactive galaxies in the Extended
Groth Strip at redshifts $0.2<z<1.1$. In particular, their figure 1
shows the increase in the SFR with redshift, and our results fall
roughly near the median SFR for similar redshift and stellar mass
(Capak et al.\ 2007) bins. This result suggests that the AGN radio
correction employed above is at the appropriate level, and that these
AGN host galaxies are forming stars at rates similar to both active
and inactive galaxies of the same size and age. This is further
evidence that host galaxies of the Compton thin $z < 1$ Seyferts that
dominate the XRB are not in the throes of a massive merger.

Interestingly, none of the 24~$\mu$m-undetected samples produce a
positive radio detection in the stacked image, which indicates that,
on average, these are very weak star-forming galaxies. To check
  for possible faint star formation among the 24~$\mu$m-undetected
  sources we make use of the observed $R-I$ and $I-K$ colors. At the
  redshifts of our sample ($z\sim0.7$) these colors approximately
  correspond to the rest-frame colors used by Williams et al.\ (2009)
  to separate faint or dusty galaxies from quiescent galaxies. Fifty
  of the 58 AGNs not detected at 24~$\mu$m have colors suggestive of
  faint star formation [$(R-I) < 0.96(I-K) + 0.4$]. If we add these
  AGNs to the samples of 24~$\mu$m-detected AGNs, and repeat the radio
  stacking measurements, the resulting SFRs are consistent with the
  SFRs presented in Sect.~\ref{proc:stacks}, which are based on the
  simple separation between detection or non-detection at
  24~$\mu$m. Thus, the majority do seem to have very little ongoing
  star formation, despite hosting a reasonably luminous AGN.
Recall that these galaxies also seem to arise from a different
population than the 24~$\mu$m-detected objects (see
Section~\ref{data:ir}), with a larger fraction of obscured,
low-luminosity AGNs. The X-ray obscuration must arise from either a
dust poor medium, or, perhaps more likely, from galactic scale
obscuration (e.g., Rigby et al.\ 2006). These AGNs are apparently
residing in relatively inactive galaxies, and thus may be in an
interesting evolutionary stage. A morphological and spectral energy
distribution study of these 24~$\mu$m-undetected AGNs is necessary to fully
understand the nature of these objects.  

In conclusion, the radio stacking results presented here, combined
with the SFRs found in the radio-detected AGNs (see Appendix A), show
that SFRs $\lesssim 40$~$M_{\odot}$~yr$^{-1}$ are present in the host
galaxies of Compton thin X-ray selected AGNs at $z < 1$. The
association of 24~$\mu$m detections with low levels of star formation
is consistent with models of nuclear starburst disks as sources for
AGN fueling and obscuration in $z < 1$ Seyfert galaxies. However, a
significant improvement in the sensitivity and resolution of radio
images for galaxies is needed to reliably determine the location of
the potential star formation that has been detected. It is possible
that future studies of radio spectral index and fractional
polarization measurements may be useful discriminants between
AGN-related and SF-related radio emission in these galaxies. Other
future work will involve examining the \textit{Herschel} photometry of
the samples, as well as a thorough investigation of the properties of
the 58 24~$\mu$m-undetected AGNs. For the present time, we can say
with certainty that Compton thin, X-ray selected AGN host galaxies at
$z < 1$ are emitting faint radio fluxes, but it is not yet clear if
the emission originates in the nuclear regions or instead in the bulge
or disk.

\acknowledgements
This work was supported in part by NSF award AST 1008067 to
D.\ R.\ B. This work is based in part on observations with the {\it
  XMM-Newton}, an ESA science mission with instruments and
contributions directly funded by ESA Member States and NASA; the
Canada-France-Hawaii Telescope with MegaPrime/MegaCam operated as a
joint project by the CFHT Corporation, CEA/DAPNIA, the NRC and CADC of
Canada, the CNRS of France, TERAPIX, and the University of Hawaii; the
Subaru Telescope, which is operated by the National Astronomical
Observatory of Japan; Kitt Peak National Observatory, Cerro Tololo
Inter-American Observatory, and the National Optical Astronomy
Observatory, which are operated by AURA, Inc., under cooperative
agreement with the National Science Foundation; and the {\it Spitzer
  Space Telescope}, which is operated by the Jet Propulsion
Laboratory, California Institute of Technology, under a contract with
NASA.

\appendix
\section{Radio-detected AGNs}\label{radio}
In Section~\ref{data:radio}, we found that 120 of the 481 X-ray selected AGNs with redshifts $z<1$ are identified as radio sources. The radio emission of some fraction of these is expected to be dominated by the AGN (often referred to as ``radio-loud''), while star formation is expected to contribute a significant portion of the radio emission from the remaining radio-detected AGNs (often known as ``radio-quiet''). Terashima \& Wilson (2003) found that $\log(R_{x}) = -3$ represents an approximate division between radio-loud [$\log(R_{x}) > -3$] and radio-quiet [$\log(R_{x}) < -3$] AGNs. Again assuming $\alpha=0.7$ for the AGN radio spectrum, we identify 24 radio-loud AGNs, which is 20\% (24/120) of the radio-detected sample and only 5\% (24/481) of the full X-ray selected AGN sample. If we instead use $\alpha=0$, we identify 31 radio-loud AGNs (26\% of the radio-detected sample), and if we use $\alpha=1$, 21 (18\%) of the radio-detected sources are identified as radio-loud. For the remainder of this Appendix, we turn our focus to the 96 radio-quiet AGNs identified using $\alpha=0.7$ and compare them to the radio-undetected AGN population described in Section~\ref{data:nh_lx}.

\begin{figure}
\plotone{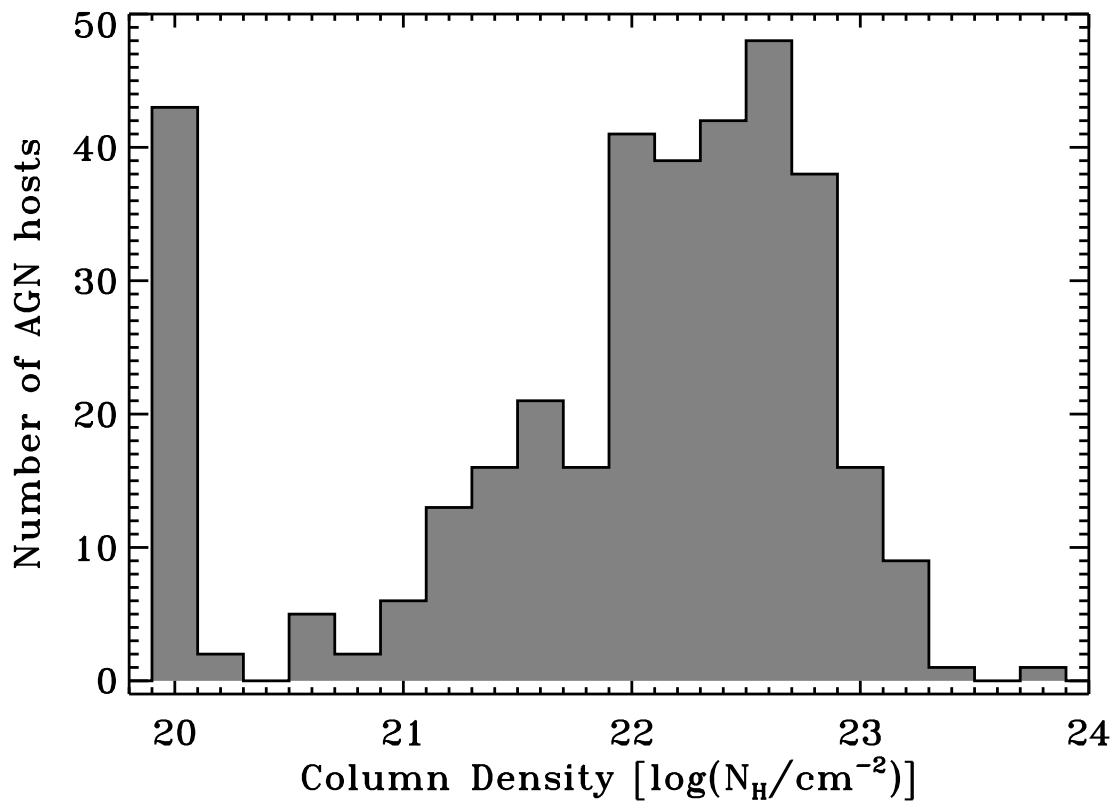}
\caption{Distribution of column densities for the full AGN sample of
  359 sources, which has a median column density
  $\log(N_{\mathrm{H}}/$cm$^{-2})=22.2$. Column densities of
  $\log(N_{\mathrm{H}}/$cm$^{-2})=20$ indicate an upper limit to the column
  densities of AGNs that experience negligible attenuation of the
  soft-band flux. A photon index of $\Gamma=1.9$ has been assumed for
  all objects.} 
\label{fig:nh}
\end{figure}

\begin{figure}
\includegraphics{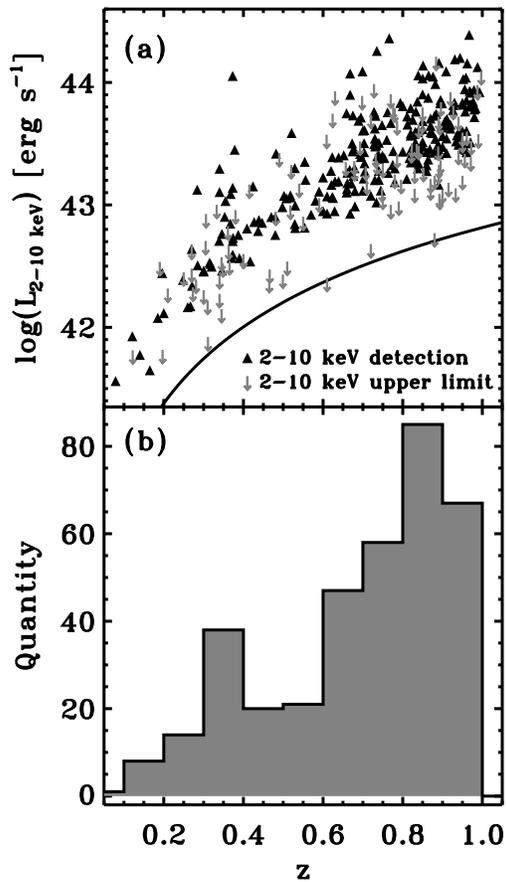}
\caption{({\it a\/}): X-ray luminosity vs.\ redshift for the AGN sample,
  represented by triangles. X-ray luminosities calculated from
  hard-band flux upper limits are indicated by gray arrows. The curve
  shows the faintest hard-band sensitivity limit (see table 2 from
  Cappelluti et al.\ 2009). ({\it b\/}): Redshift distribution of the
  full AGN sample.} 
\label{fig:lx_z}
\end{figure}

\begin{figure}
\plotone{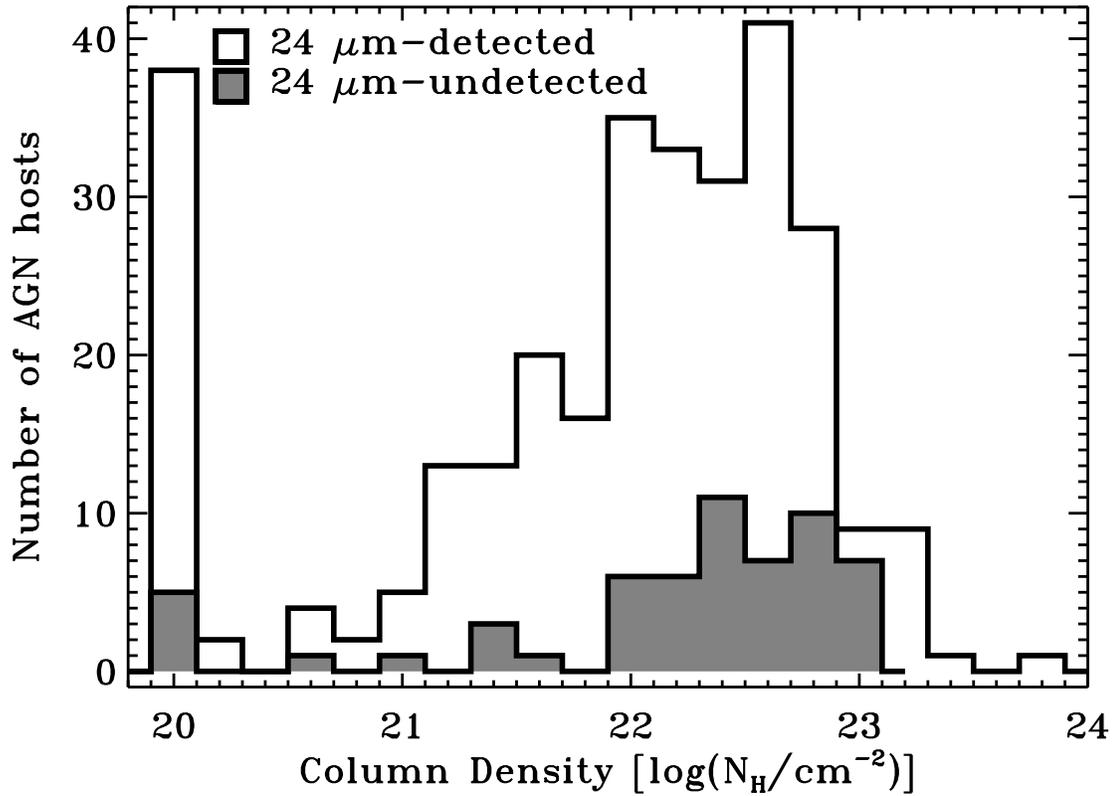}
\caption{Column density distributions of the 24~$\mu$m-detected (open
  black histogram; 301 AGNs) and undetected (filled gray histogram; 58
  AGNs) AGNs. According to a K-S test, the distributions are
  statistically different (see text for details).} 
\label{fig:nh_ir}
\end{figure}

\begin{figure}
\includegraphics{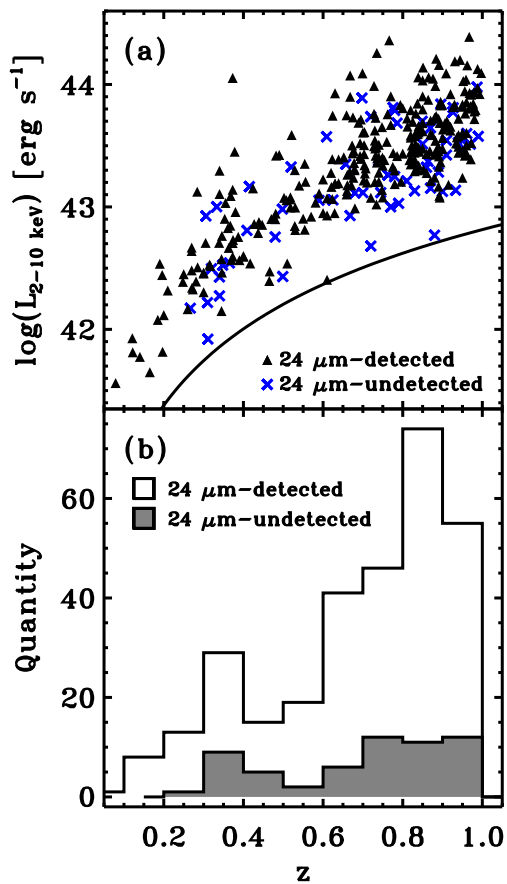}
\caption{({\it a\/}): X-ray luminosity vs.\ redshift for the
  24~$\mu$m-detected (black triangles) and undetected (blue crosses)
  AGNs. The curve is as in Figure~\ref{fig:lx_z}. ({\it b\/}): Redshift
  distributions of the 24~$\mu$m-detected and undetected AGNs, as in
  Figure~\ref{fig:nh_ir}. Statistically, a K-S test shows that the
  redshift distributions are very similar, but the X-ray luminosity
  distributions differ (see text for details).} 
\label{fig:lx_z_ir}
\end{figure}

\begin{figure}
\includegraphics[width=0.33\textwidth]{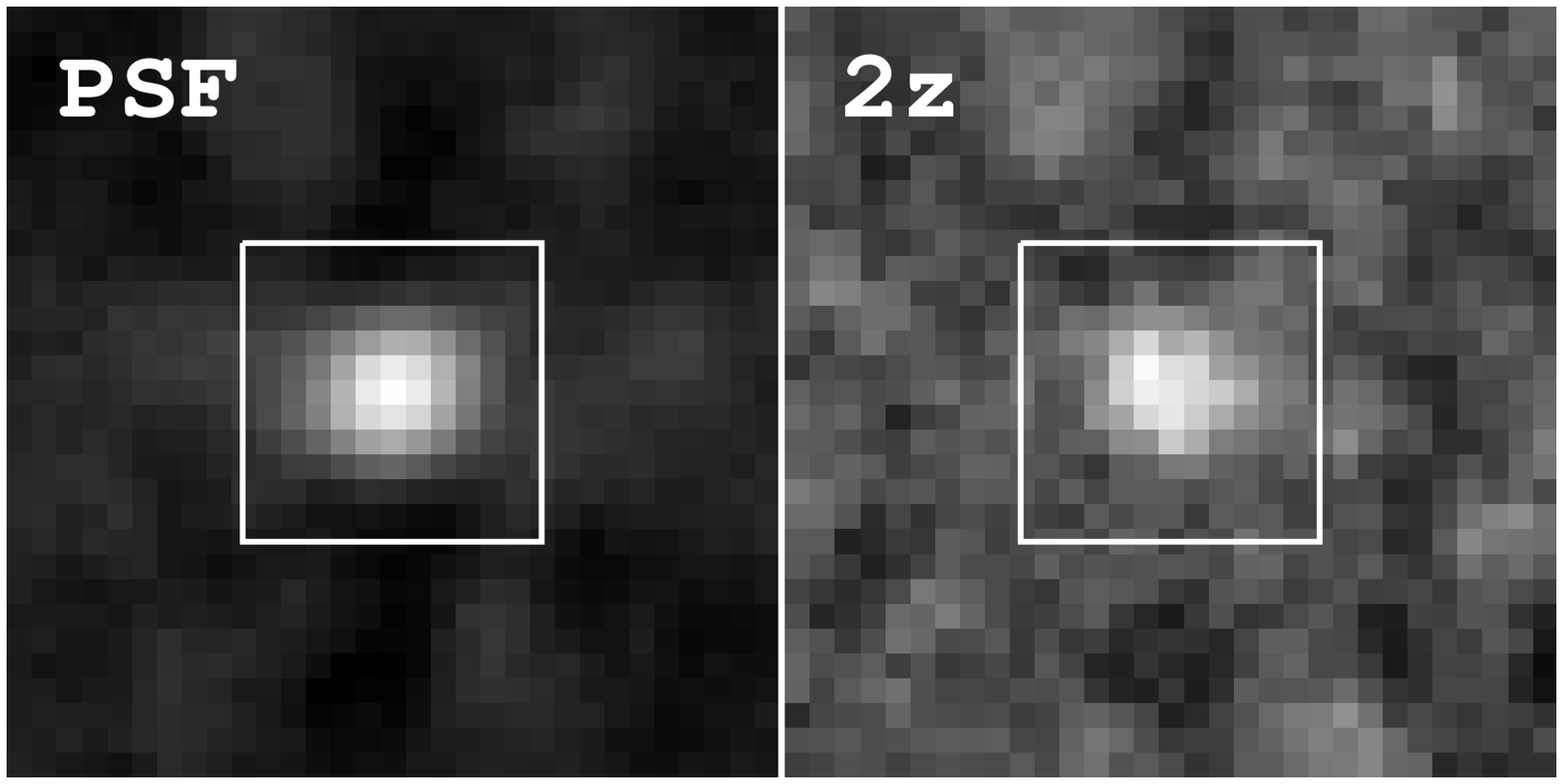}\\
\includegraphics[width=0.5\textwidth]{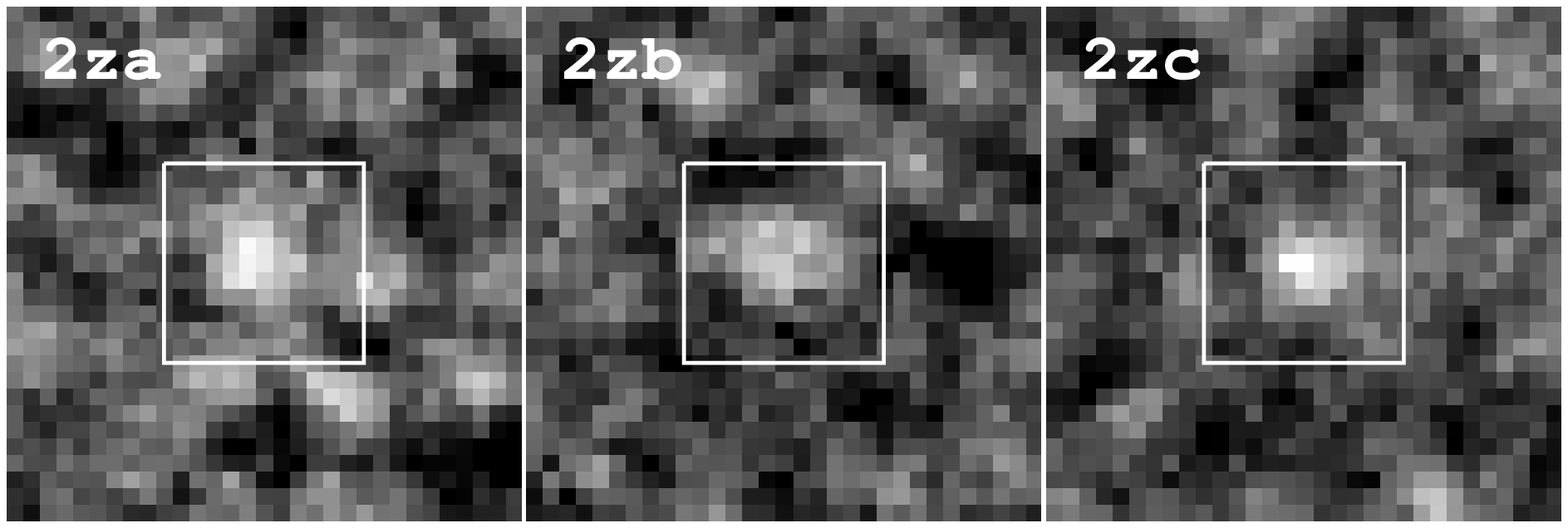}\\
\includegraphics[width=0.5\textwidth]{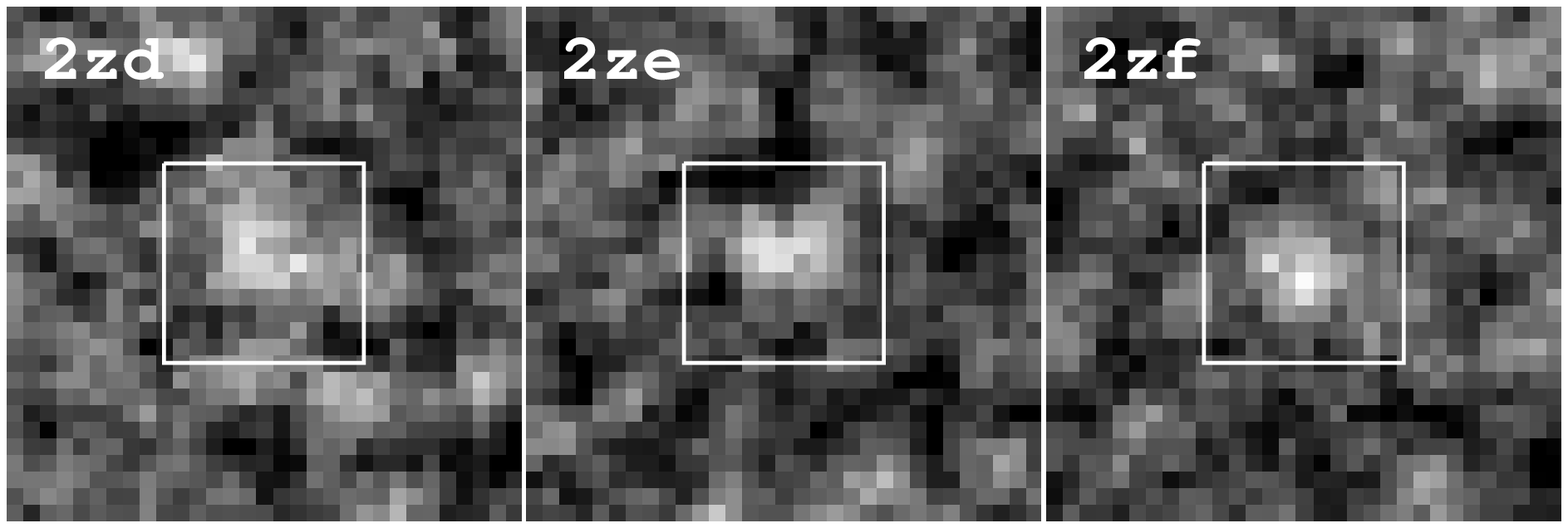}\\
\includegraphics[width=0.5\textwidth]{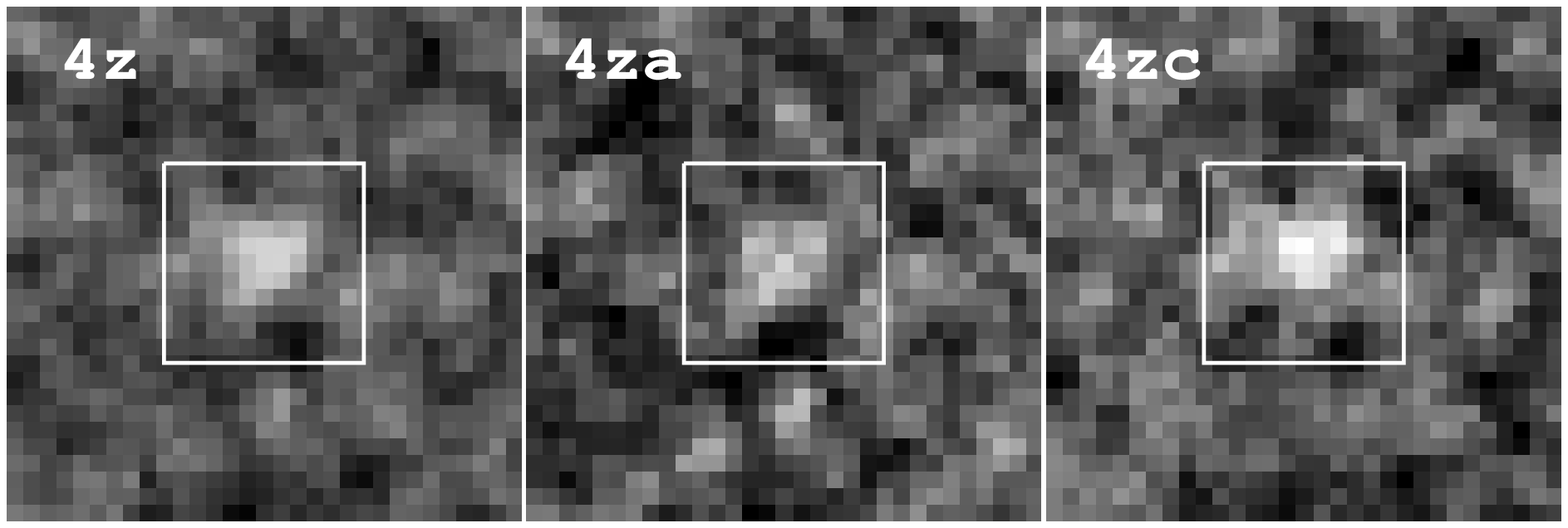}\\
\includegraphics[width=0.5\textwidth]{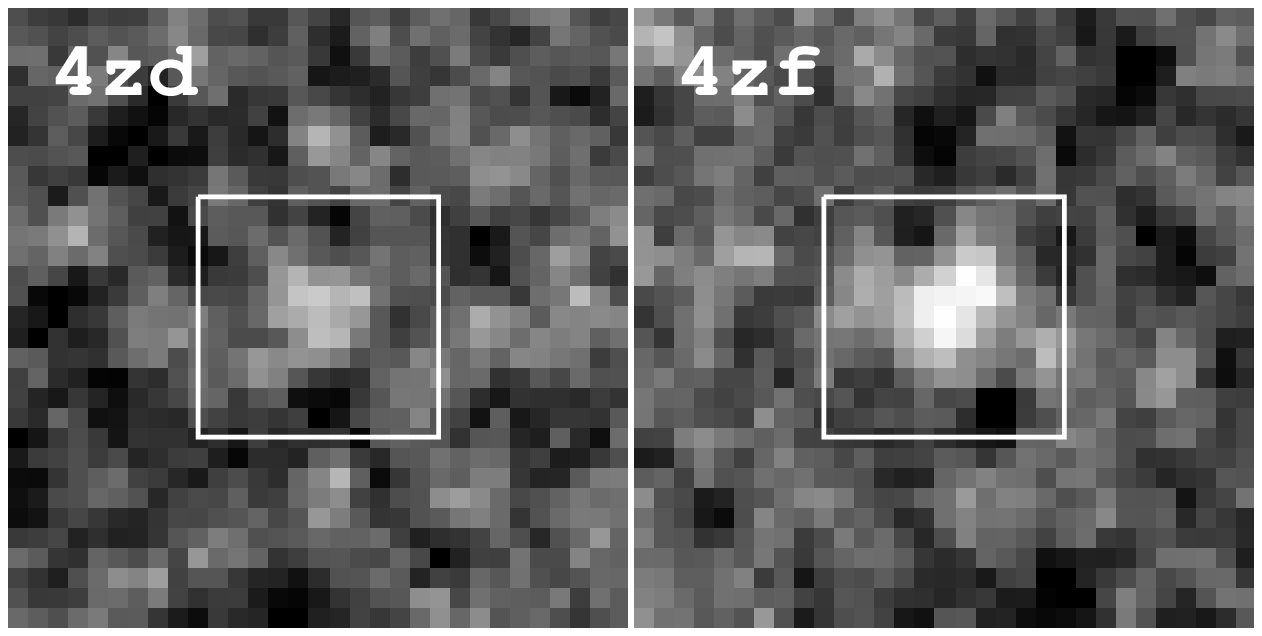}\\
\includegraphics[width=0.5\textwidth]{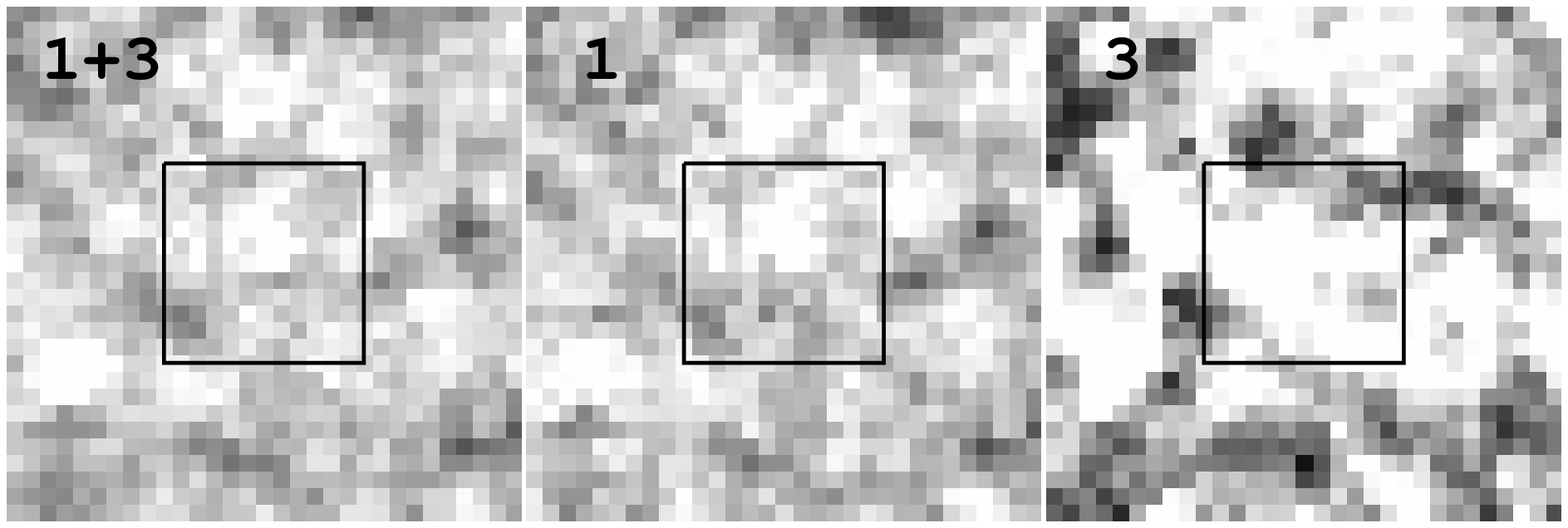}
\caption{Stacked residual radio images. The boxes denote the 12 pixel
  x 12 pixel (4.2\arcsec\ x 4.2\arcsec) regions used to determine the
  integrated radio flux densities (Section~\ref{proc:aips}). All of the sample
  images were created using the same linear scaling. The color scale of the
  final row has been inverted
  to better show the details. The scaling for the PSF image is unique
  to better show the image. Each panel, accept for the PSF, represents
  one of the samples listed in Table~\ref{table:samples}, as labeled.}
\label{fig:stamps}
\end{figure}

\begin{figure}
\includegraphics[width=0.75\textwidth]{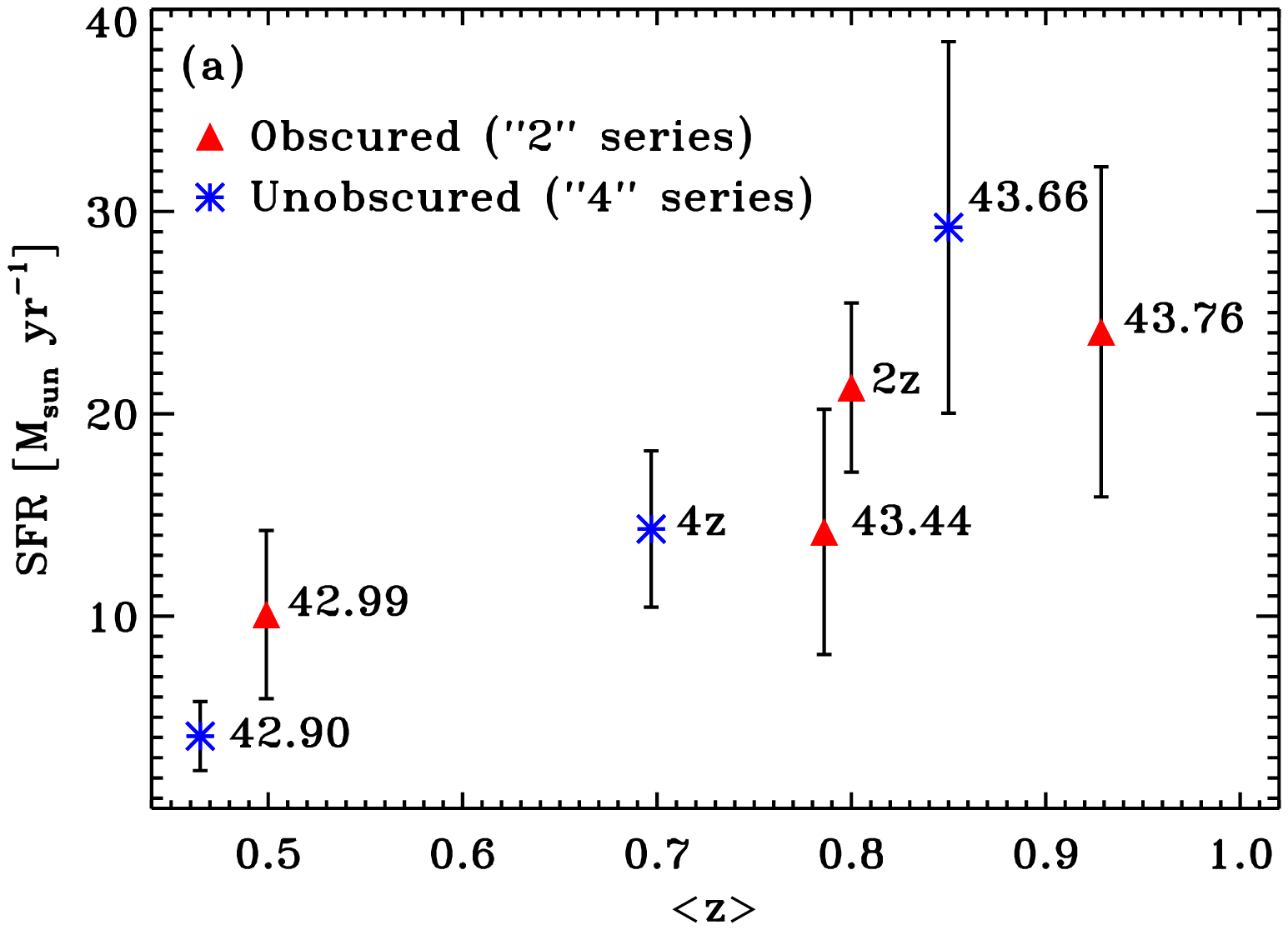}\\
\includegraphics[width=0.75\textwidth]{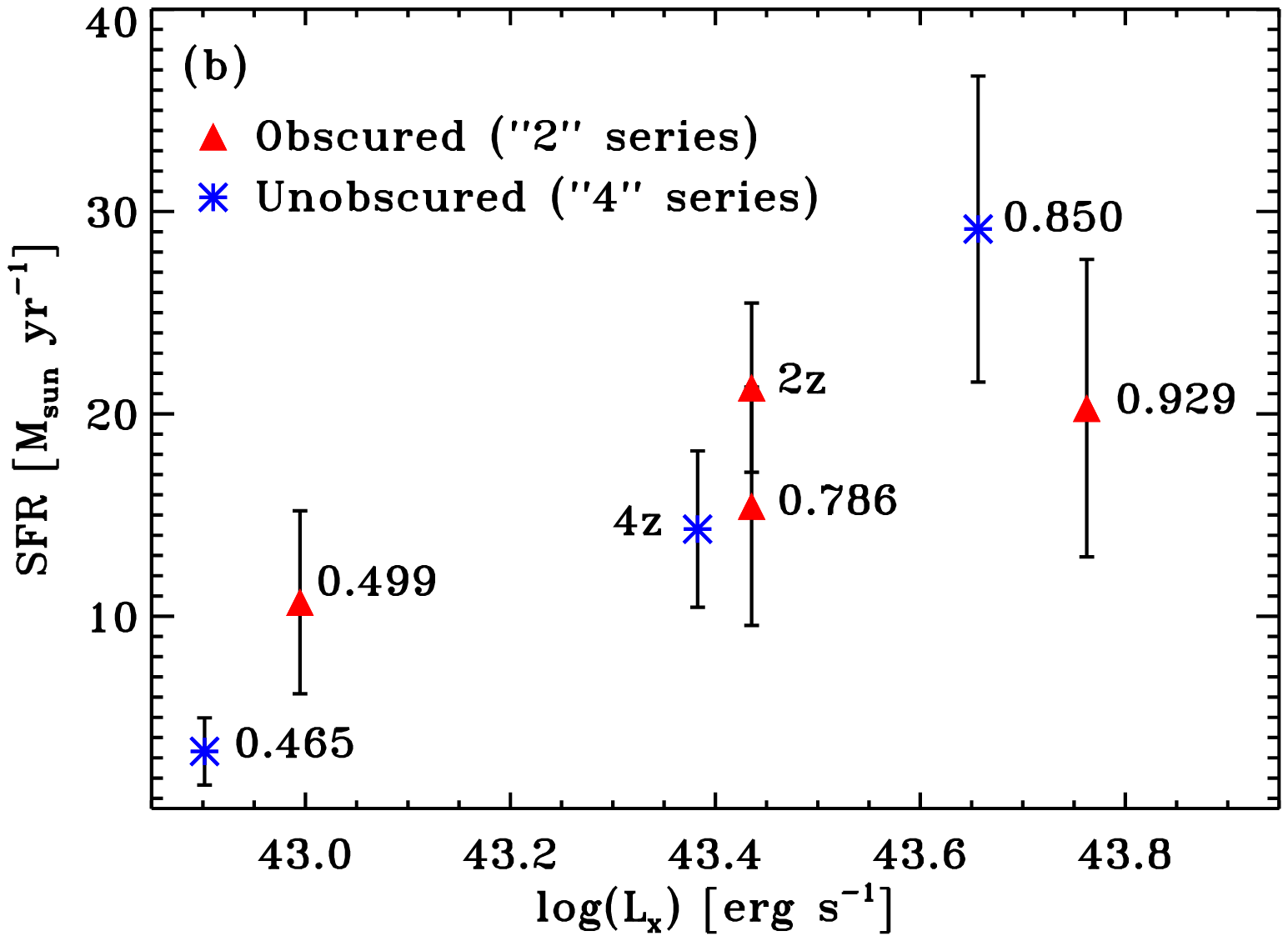}
\caption{Star formation rate (SFR) as a function of redshift (upper panel)
  and X-ray luminosity (lower panel). Red triangles represent obscured
  AGN samples, and blue stars represent unobscured AGN samples. The
  symbols marked ``2z'' and ``4z'' represent the full obscured and
  unobscured 24~$\mu$m-detected samples, respectively. The other
  symbols are marked with the median luminosity (upper panel) or
  redshift (lower panel) of the sample. Error bars
  indicate the 1$\sigma$ uncertainties on the SFRs,
  estimated from the uncertainties on the measured integrated flux densities.}
\label{fig:sfr}
\end{figure}

\begin{figure}
\plotone{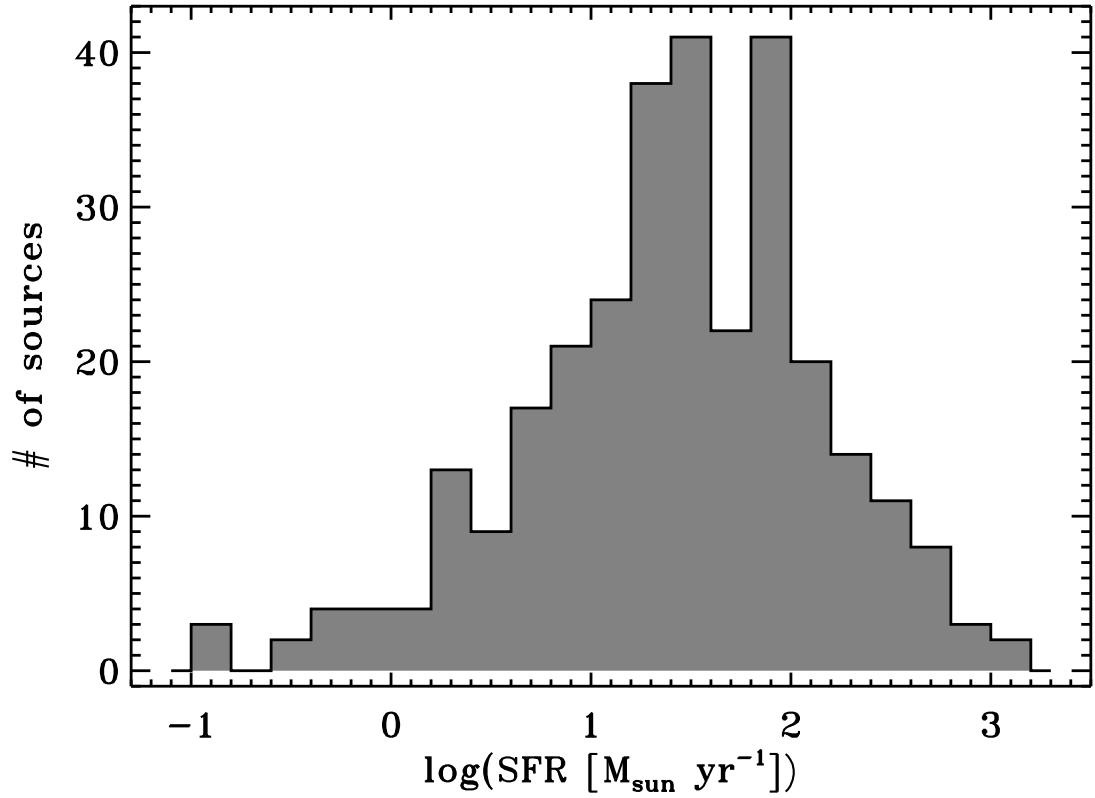}
\caption{Distribution of the 24~$\mu$m derived star formation rates (SFRs)
  for the 301 radio undetected AGNs with a 24~$\mu$m detection. The
  SFRs were calculated using the Rieke et al.\ (2009) calibration. The
  median SFR is 29~$\Msun$~yr$^{-1}$ which is equal to the maximum SFR that
  is found from the stacked radio flux densities.} 
\label{fig:ir_flux}
\end{figure}

\begin{figure}
\figurenum{A1}
\plotone{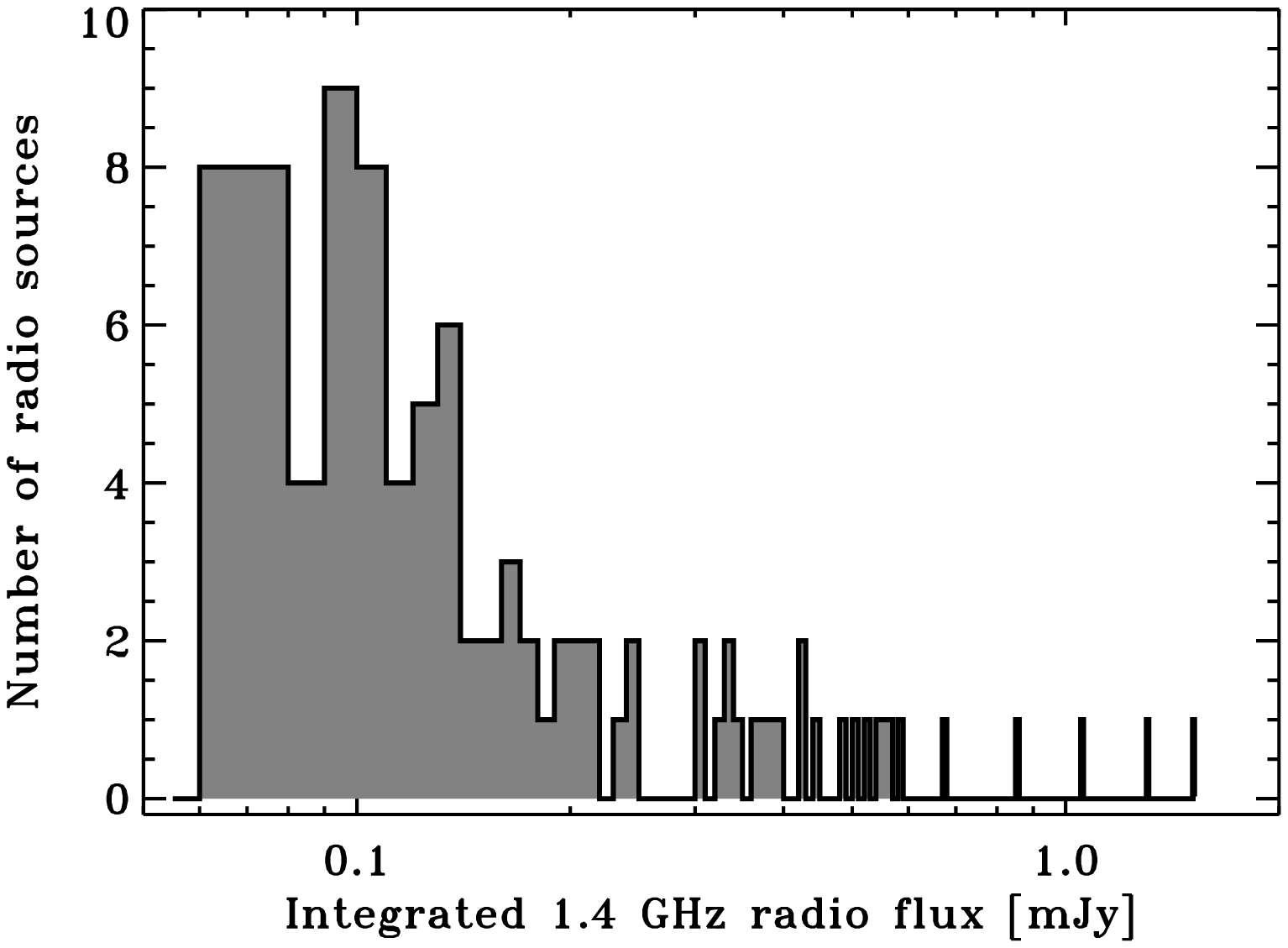}
\caption{Radio flux density distribution of the radio-quiet AGNs at redshifts
  $z<1$. The radio emission from these systems might be expected to be dominated by star formation.}
\label{fig:radio_flux_agn}
\end{figure}

\begin{figure}
\figurenum{A2}
\plotone{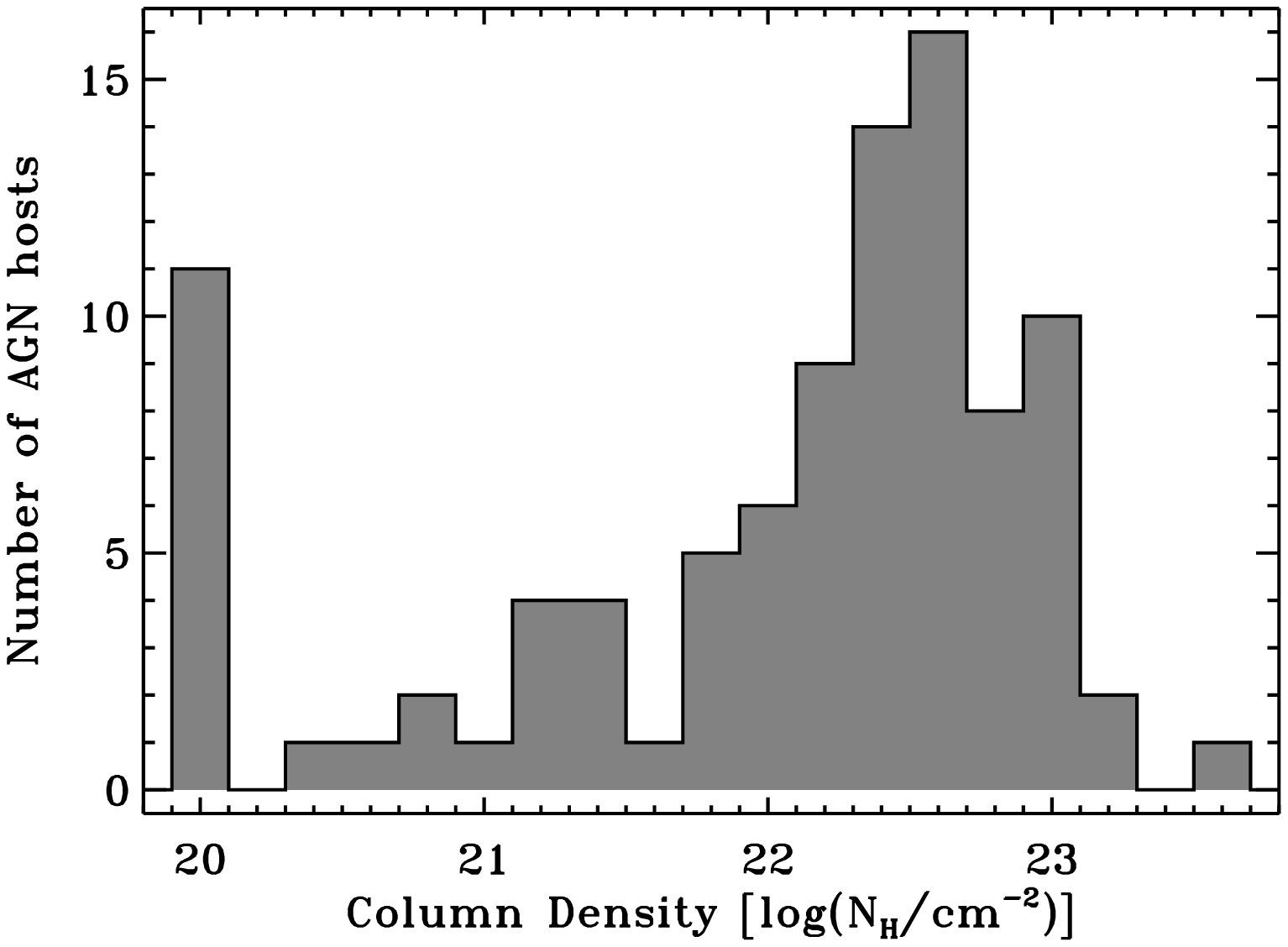}
\caption{Column density distribution of the radio-quiet AGNs at redshifts $z<1$.}
\label{fig:nh_radio}
\end{figure}

\begin{figure}
\figurenum{A3}
\includegraphics{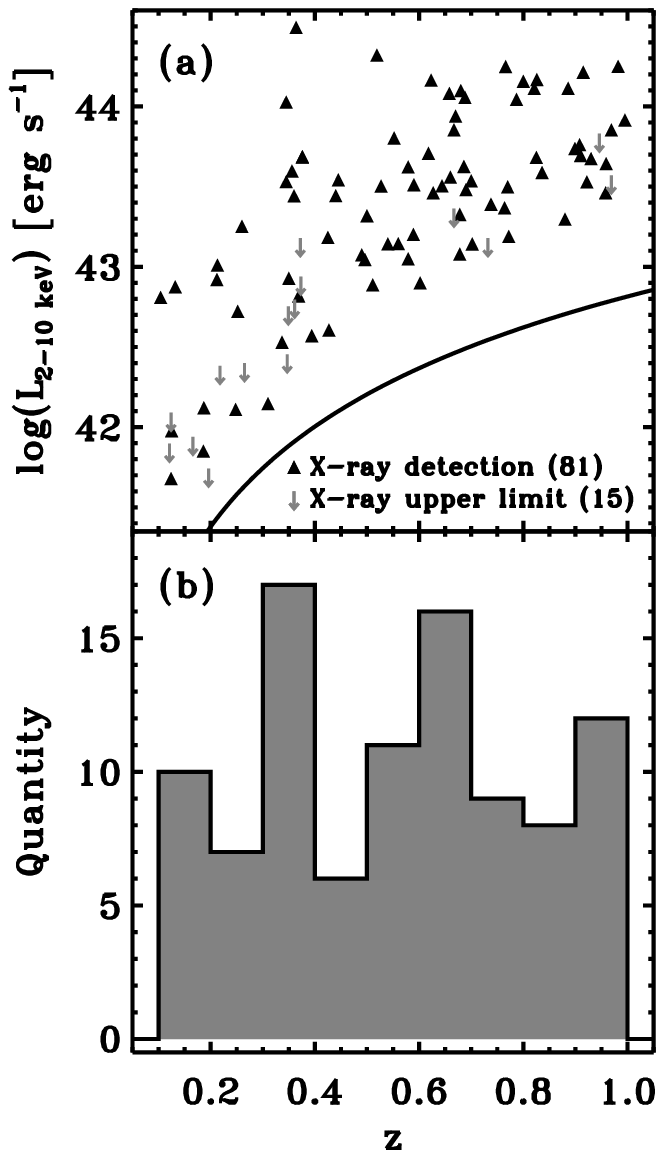}
\caption{({\it a\/}): X-ray luminosity vs.\ redshift for the radio-quiet AGNs. Symbols and the curve are as in Figure~\ref{fig:lx_z}, except that these are all radio-quiet AGNs. ({\it b\/}): Redshift distribution of the radio-quiet AGNs.}
\label{fig:lx_z_radio}
\end{figure}

The distribution of radio flux densities for the radio-quiet AGNs is shown in Figure~\ref{fig:radio_flux_agn}. These systems have a median radio flux density of $f = 0.133$~mJy. Figure~\ref{fig:nh_radio} shows the column density distribution of the 96 radio-quiet AGNs, which can be compared to Figure~\ref{fig:nh}. The median column density is $\log(N_{\mathrm{H}}/$cm$^{-2}) = 22.4$, and a K-S test between the column densities of the radio-quiet AGNs and the radio-undetected AGNs confirms the apparent similarity between Figures~\ref{fig:nh} and \ref{fig:nh_radio}, with a 23\% probability of having been drawn from the same parent population and a low maximum deviation ($D=0.12$). The X-ray luminosities and redshift distribution of the radio-quiet AGNs are shown in Figure~\ref{fig:lx_z_radio} (compare to Figure~\ref{fig:lx_z}). The radio-quiet AGNs exhibit a broader X-ray luminosity distribution than the radio-undetected AGNs, especially at redshifts $z<0.6$. Statistically, however, the X-ray luminosity distributions are not very different ($D=0.098$, $p=0.44$). In contrast, panel (b) of Figure~\ref{fig:lx_z_radio} shows that the radio-detected AGNs have a much flatter redshift distribution than do the radio undetected AGNs. A K-S test confirms this difference; the maximum deviation is $D=0.30$ and the probability that the distributions represent the same sample is quite low ($p<<1$). Finally, $89$ of the $96$ radio-quiet AGNs are also 24~$\mu$m-detected.

\begin{figure}
\figurenum{A4}
\includegraphics{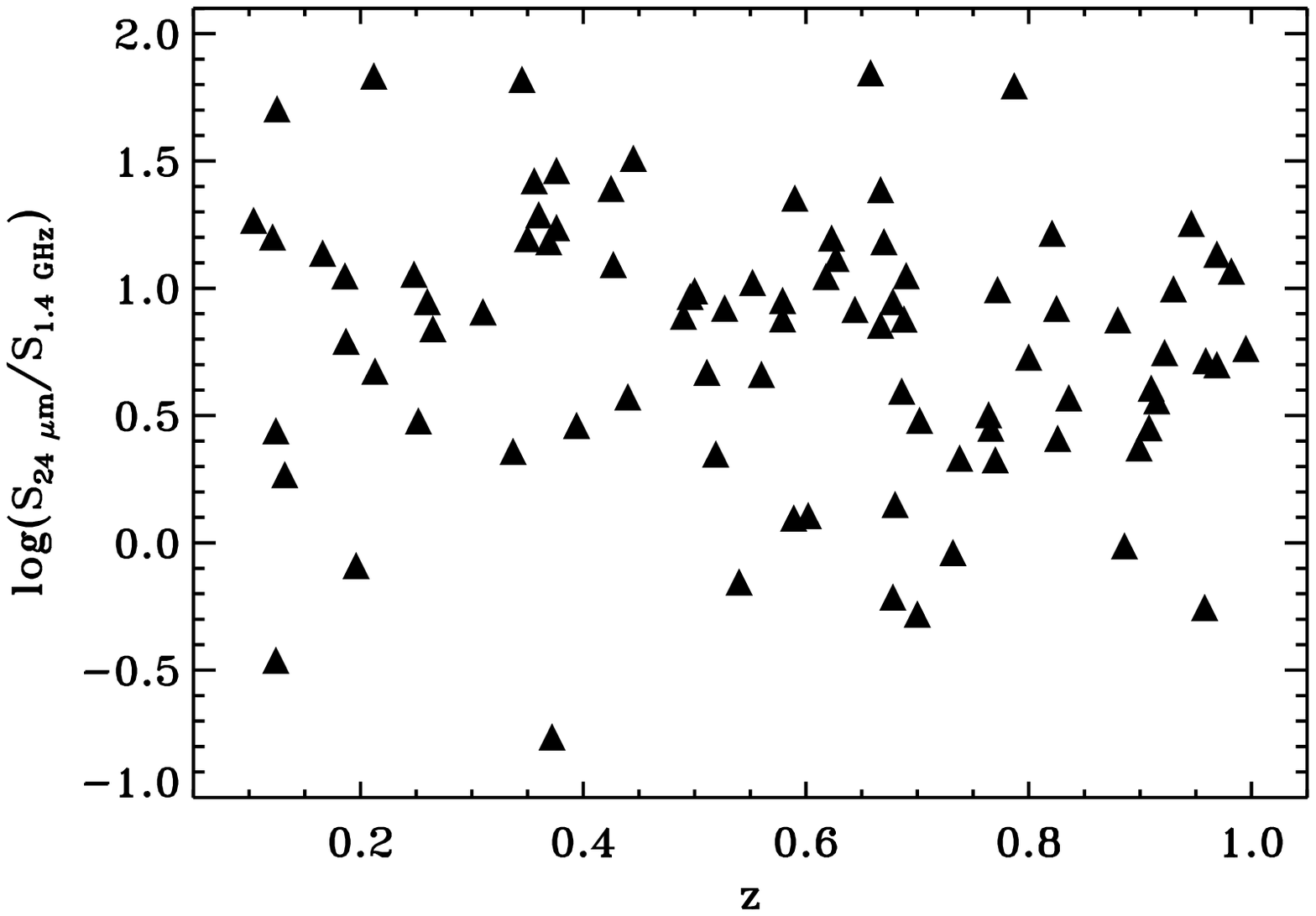}
\label{fig:seymour}
\caption{Ratio of 24~$\mu$m to 1.4~GHz flux density for the 89 radio detected
  $z<1$ AGNs with 24~$\mu$m detections. The radio flux density has been
  corrected for the nuclear AGN emission. The median value is $0.88$,
  which clearly indicates that the radio emission is dominated by star
  formation (Seymour et al.\ 2008).}
\end{figure}

Eqs.~\ref{eq:rx}--\ref{eq:flux} are again used to estimate and remove the AGN contribution to the measured radio flux density. As before, we use a spectral index $\alpha=0.7$, but we also check the effect of varying the spectral index. The median predicted 1.4~GHz flux density for the radio-quiet sources is $f=12~\mu$Jy when we use $\alpha=0.7$. A flatter spectrum ($\alpha=0$) results in a lower median predicted flux density ($f=6.4~\mu$Jy). Using a steeper spectrum ($\alpha=1$) increases the predicted flux density to $f=15~\mu$Jy. Figure~\ref{fig:seymour} plots the mid-IR to residual radio flux density ratio for those AGNs detected at 24~$\mu$m. The logarithm of the median flux ratio is $0.88$, which strongly supports the interpretation that the residual radio emission originates from star formation (Seymour et al.\ 2008). 

\begin{figure}
\figurenum{A5}
\plotone{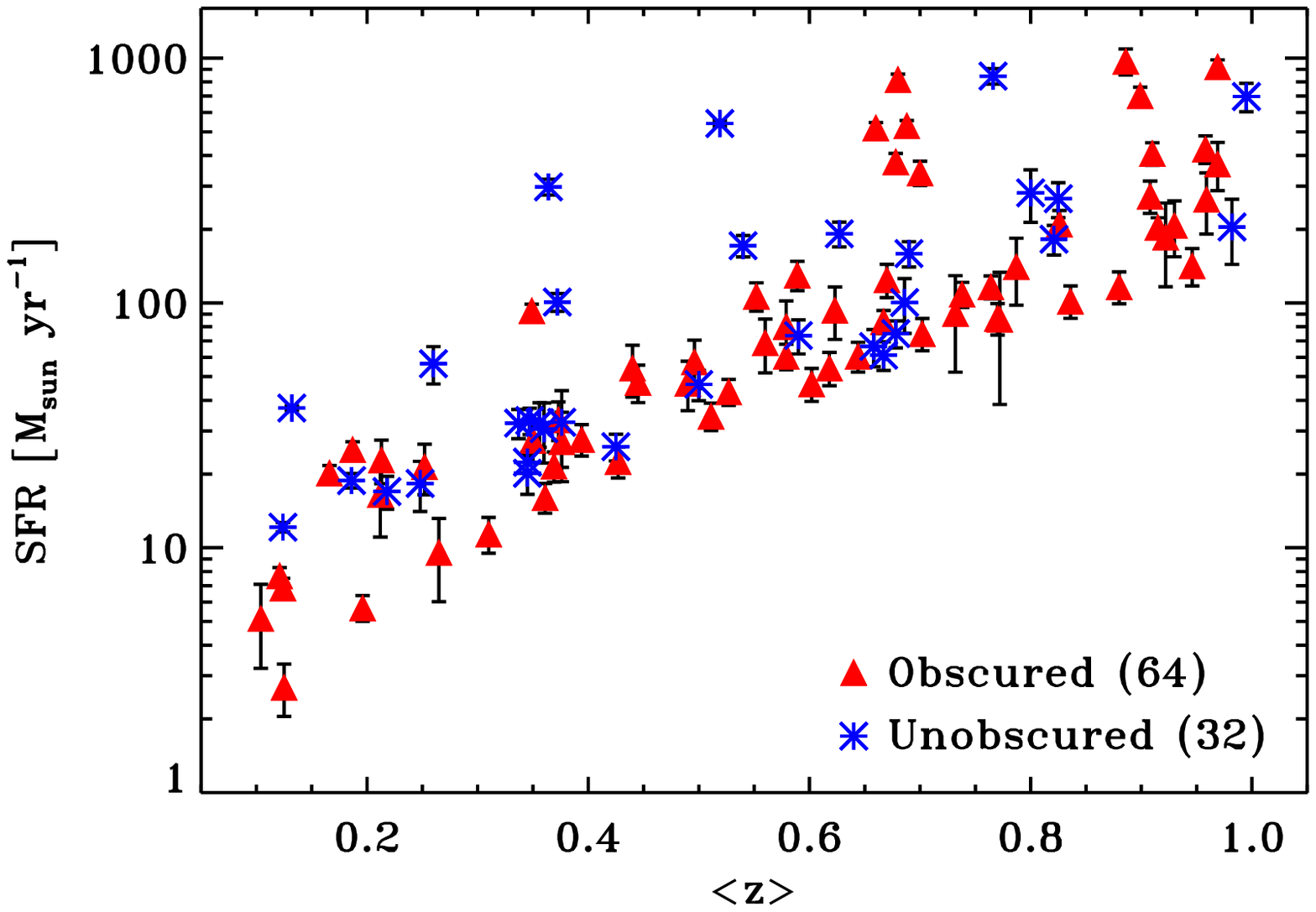}
\caption{Star formation rates as a function of redshift for the
  obscured (red triangles) and unobscured (blue stars) radio-quiet
  AGNs. Error bars represent the uncertainties from the integrated
  radio flux density measurements.}
\label{sfr_z_radio}
\end{figure}

Figure~\ref{sfr_z_radio} shows the SFRs computed from the residual radio flux densities as a function of redshift for the obscured and unobscured radio-quiet AGNs, with median rates of 77~$\Msun$~yr$^{-1}$ and 42~$\Msun$~yr$^{-1}$, respectively, and an overall median rate of 65~$\Msun$~yr$^{-1}$. Increasing the spectral index to $\alpha=1$ increases the median SFRs to 72~$\Msun$~yr$^{-1}$, and decreasing it to $\alpha=0$ similarly decreases the median star formation rate to 50~$\Msun$~yr$^{-1}$.

More than 90\% of the radio-quiet AGN host galaxies are undergoing star formation at rates in excess of 10~$\Msun$~yr$^{-1}$, and 35\% are experiencing SFRs exceeding 100~$\Msun$~yr$^{-1}$. These values are again largely consistent with those predicted by nuclear starburst disks (Ballantyne 2008), although with a larger fraction of high SFRs ($>100$~$\Msun$~yr$^{-1}$). This fact will be largely due to the increase in detectable SFRs with $z$ in a flux limited survey. Combining this result with the stacking analysis from Section~\ref{proc:stacks}, indicates that the majority of 24~$\mu$m-detected, X-ray selected AGNs at $z < 1$ are accompanied by star formation at levels of $\lesssim 40$~$\Msun$~yr$^{-1}$.

\references{}
\q
Antonucci, R.\ 1993, ARA\&A, 31, 473
\q
Adelman-McCarthy, J.\ K., et al.\ 2006, ApJS, 162, 38
\q
Ballantyne, D.\ R., Everett, J.\ E., \& Murray, N.\ 2006, ApJ, 639, 740
\q
Ballantyne, D.\ R.\ 2008, ApJ, 685, 787
\q
---.\ 2009, ApJ, 698, 1033
\q
Bauer, F.\ E., Alexander, D.\ M., Brandt, W.\ N., Hornschemeier, A.\ E., Vignali, C., Gamire, G.\ P., \& Schneider, D.\ P.\ 2002, AJ, 124, 1839
\q
Bell, E.\ F.\ 2003, ApJ, 586, 794
\q
Biggs, A.\ D., \& Ivison, R.\ J.\ 2006, MNRAS, 371, 963
\q
Bondi, M., Ciliegi, P., Schinnerer, E., Smol$\check{\rm c}$i\'{c}, V., Jahnke, K., Carilli, C., \& Zamorani, G.\ 2008, ApJ, 681, 1129
\q
Boyle, B.\ J., Cornwell, T.\ J., Middelberg, E., Norris, R.\ P., Appleton, P.\ N., \& Smail, I.\ 2007, MNRAS, 376, 1182
\q
Brand, K., et al.\ 2006, ApJ, 644, 143
\q
Brusa, M., et al.\ 2010, ApJ, 716, 348
\q
Capak, P., et al.\ 2007, ApJS, 172, 116
\q
Cappelluti, N., et al.\ 2009, A\&A, 497, 635
\q
Carilli, C.\ L., et al.\ 2008, ApJ, 689, 883
\q
Chang, P., Quataert, E., \& Murray, N.\ 2007, ApJ, 662, 94
\q
Chary, R., \& Elbaz, D.\ 2001, ApJ, 556, 562
\q
de Vries, W.\ H., Hodge, J.\ A., Becker, R.\ H., White, R.\ L., \& Helfand, D.\ J.\ 2007, AJ, 134, 457
\q
Draper, A.\ R., \& Ballantyne, D.\ R.\ 2010, ApJ, 715, L99
\q
Dunne, L., et al.\ 2009, MNRAS, 394, 3
\q
Elvis, M., et al.\ 2009, ApJS, 184, 158
\q
Elitzur, M., \& Shlosman, I.\ 2006, ApJ, 648, L101
\q
Fabian, A.\ C., Barcons, X., Almaini, O., \& Iwasawa, K.\ 1998, MNRAS, 297, L11
\q
Fasano, G., \& Franceschini, A.\ 1987, MNRAS, 225, 155
\q
Gilli, R., Comastri, A., \& Hasinger, G.\ 2007, A\&A, 463, 79
\q
Hasinger, G.\ 2008, A\&A, 490, 905
\q
Hasinger, G., et al.\ 2007, ApJS, 172, 29
\q
Hatziminaoglou, E., Fritz, J., \& Jarrett, T.\ H., 2009, MNRAS, 399, 1206
\q
Ho, L.\ C., Filippenko, A.\ V., \& Sargent, W.\ L.\ W.\ 1997, ApJ, 487, 568
\q
Ho, L.\ C., \& Ulvestad, J.\ S.\ 2001, ApJS, 133, 77
\q
Hodge, J.\ A., Becker, R.\ H., White, R.\ L., \& de Vries, W.\ H.\ 2008, AJ, 136, 1097
\q
Ibar, E., \& Lira, P., 2007, A\&A, 466, 531
\q
Ivison, R.\ J., et al.\ 2007, ApJ, 660, L77
\q
Karim, A., et al.\ 2011, ApJ, 730, 61
\q
Kauffmann, G., et al.\ 2003, MNRAS, 346, 1055
\q
K$\ddot{\rm o}$nigl, A., \& Kartje, J.\ F.\ 1994, ApJ, 434, 446
\q
Krolik, J.\ H.\ 2007, ApJ, 661, 52
\q
Krolik, J.\ H., \& Begelman, M.\ C.\ 1988, ApJ, 329, 702
\q
Le Floc'h, E., et al.\ 2005, ApJ, 632, 169
\q
---.\ 2009, ApJ, 703, 222
\q
Lilly, S.\ J., et al.\ 2007, ApJS, 172, 70
\q
---.\ 2008, Messenger, 134, 35
\q
---.\ 2009, ApJS, 184, 218
\q
Lutz, D., et al.\ 2010, ApJ, 712, 1287
\q
Mainieri, V., et al.\ 2007, ApJS, 172, 368
\q
Maiolino, R., \& Rieke, G.\ H.\ 1995, ApJ, 454, 95
\q
Malizia, A., Stephen, J.\ B., Bassani, L., Bird, A.\ J., Panessa, F., \& Ubertini, P.\ 2009, MNRAS, 399, 944
\q
Morrison, G.\ E., Owen, F.\ N., Dickinson, M., Ivison, R.\ J., \& Ibar, E.\ 2010, ApJS, 188, 178
\q
Mullaney, J.\ R., et al.\ 2011, MNRAS, submitted, arXiv:1106.4284v2
\q
Nagar, N.\ M., Falcke, H., Wilson, A.\ S., \& Ulvestad, J.\ S.\ 2002, A\&A, 392, 53
\q
Nandra, K., \& Pounds, K.\ A.\ 1994, MNRAS, 268, 405
\q
Nenkova, M., Sirocky, M.\ M., Ivezi\'{c}, \v{Z}., \& Elitzur, M.\ 2008a, ApJ, 685, 147
\q 
Nenkova, M., Sirocky, M.\ M., Nikutta, R., Ivezi\'{c}, \v{Z}., \& Elitzur, M.\ 2008b, ApJ, 685, 160
\q
Noeske, K.\ G., et al.\ 2007, ApJ, 660, L43
\q
Panessa, F., Barcons, X., Bassani, L., Cappi, M., Carrera, F.\ J., Ho, L.\ C., \& Pellegrini, S.\ 2007, A\&A, 467, 519
\q
Park, S.\ Q., et al.\ 2008, ApJ, 678, 744
\q
Pier, E.\ A., \& Krolik, J.\ H.\ 1992, ApJ, 401, 99
\q
Prescott, M.\ K.\ M., Impey, C.\ D., Cool, R.\ J., \& Scoville, N.\ Z.\ 2006, ApJ, 644, 100
\q
Richards, E.\ A.\ 2000, ApJ, 533, 611
\q
Rieke G.\ H., Alonso-Herrero A., Weiner B.\ J., P\'{e}rez-Gonz\'{a}lez P.\ G., Blaylock M., Donley J.\ L., \& Marcillac D.\ 2009, ApJ, 692, 556
\q
Rigby, J.\ R., Rieke, G.\ H., Donley, J.\ L., Alonso-Herrero, A., \& P\'{e}rez-Gonz\'{a}lez, P.\ G.\ 2006, ApJ, 645, 115
\q
Sanders, D.\ B., et al.\ 2007, ApJS, 172, 86
\q
Salvato, M., et al.\ 2009, ApJ, 690, 1250
\q
Schinnerer, E., et al.\ 2004, AJ, 128, 1974
\q
---.\ 2007, ApJS, 172, 46
\q
Scoville, N., et al.\ 2007, ApJS, 172, 1
\q
Seymour, N., Dwelly, T., Moss, D., McHardy, I., Zoghbi, A., Rieke, G., Page, M., Hopkins, A., \& Loaring, N.\ 2008, MNRAS, 386, 1695
\q
Shi, J., \& Krolik, J.\ H.\ 2008, ApJ, 679, 1018
\q
Strazzullo, V., Pannella, M., Owen, F.\ N., Bender, R., Morrison, G.\ E., Wang, W.-H., \& Shupe, D.\ L.\ 2010, ApJ, 714, 1305
\q
Terashima, Y., \& Wilson, A.\ S.\ 2003, ApJ, 583, 145
\q
Thompson, T.\ A., Quataert, E., \& Murray, N.\ 2005, ApJ, 630, 167
\q
Treister, E., Natarajan, P., Sanders, D.\ B., Urry, C.\ M., Schawinski, K., \& Kartaltepe, J.\ 2010, Science, 328, 600
\q
Trump, J.\ R., et al.\ 2007, ApJS, 172, 383
\q
---.\ 2009, ApJ, 1195
\q
Ueda, Y., Akiyama, M., Ohta, K., \& Miyaji, T.\ 2003, ApJ, 598, 886
\q
Wada, K., \& Norman, C.\ 2002, ApJ, 566, L21
\q
Wals, M., Boyle, B.\ J., Croom, S.\ M., Miller, L., Smith, R., Shanks, T., \& Outram, P.\ 2005, MNRAS, 360, 453
\q
White, R.\ L., Helfand, D.\ J., Becker, R.\ H., Glikman, E., \& de Vries, W.\ 2007, ApJ, 654, 99
\q
Williams, R.\ J., Quadri, R.\ F., Franx, M., van Dokkum, P., \& Labb\'{e}, I.\ 2009, ApJ, 691, 1879

\clearpage

\begin{deluxetable}{lcccc}
\centering
\setlength{\tabcolsep}{0.05in}
\tablewidth{0pc}
\tablecolumns{5}
\tablecaption{COSMOS observations and data\label{table:data}}
\tablehead{
  \colhead{Telescope} & \colhead{Band} &
  \colhead{$\lambda_{\rm eff}$} & \colhead{$1\sigma$ Sensitivity Limits} & \colhead{References}
}
\startdata
{\it XMM-Newton} & Hard band &  3.1 \AA \ (4 keV) & $9.3 \times 10^{-15}$ erg s$^{-1}$ cm$^{-2}$ & (1) \\
                 & Soft band & 12.4 \AA \ (1 keV) & $1.7 \times 10^{-15}$ erg s$^{-1}$ cm$^{-2}$ & (1) \\
Very Large Array & 1.4 GHz   & 21 cm              & 11 $\mu$Jy/beam                                  & (2)
\enddata
\tablerefs{(1) Hasinger et al.\ 2007, Cappelluti et al.\ 2009. (2) Schinnerer et al.\ 2007, Bondi et al.\ 2008.}
\end{deluxetable}

\clearpage

\begin{deluxetable}{lrcccccccc}
\setlength{\tabcolsep}{0.05in}
\tablewidth{0pc}
\tablecolumns{9}
\tablecaption{AGN sub-samples\label{table:samples}}
\tablehead{
  \colhead{Sample} & \colhead{\#}  & \colhead{z$_{\rm min}$} & \colhead{z$_{\rm max}$} & \colhead{$\langle z \rangle$} & \colhead{$\log(L_{x})_{\rm min}$} & \colhead{$\log(L_{x})_{\rm max}$} & \colhead{$\langle \log(L_{x}) \rangle$} & \colhead{$\langle \log(N_{\mathrm{H}}/$cm$^{-2}) \rangle$} \\
  \colhead{(1)}    & \colhead{(2)} & \colhead{(3)}           & \colhead{(4)}           & \colhead{(5)}   & \colhead{(6)}                     & \colhead{(7)}                     & \colhead{(8)}             & \colhead{(9)}
}
\startdata
\cutinhead{24 $\mu$m-detected, $\log(N_{\mathrm{H}}/$cm$^{-2})>$ 22}
2z            & 165 & 0.000 & 1.000 & 0.800 & 41.50 & 44.50 & 43.44 & 22.52 \\
2za           &  54 & 0.000 & 0.678 & 0.499 & 41.50 & 44.50 & 42.98 & 22.37 \\
2zb           &  55 & 0.678 & 0.859 & 0.786 & 41.50 & 44.50 & 43.49 & 22.53 \\
2zc           &  56 & 0.859 & 1.000 & 0.929 & 41.50 & 44.50 & 43.66 & 22.58 \\
2zd           &  55 & 0.000 & 1.000 & 0.518 & 41.50 & 43.27 & 42.99 & 22.33 \\
2ze           &  55 & 0.000 & 1.000 & 0.830 & 43.27 & 43.59 & 43.44 & 22.52 \\
2zf           &  55 & 0.000 & 1.000 & 0.884 & 43.59 & 44.50 & 43.76 & 22.66 \\
\cutinhead{24 $\mu$m-detected, $\log(N_{\mathrm{H}}/$cm$^{-2})<$ 22}
4z            & 136 & 0.000 & 1.000 & 0.697 & 41.50 & 44.50 & 43.38 & 21.39 \\
4za           &  69 & 0.000 & 0.700 & 0.465 & 41.50 & 44.50 & 42.90 & 21.39 \\
4zc           &  67 & 0.700 & 1.000 & 0.850 & 41.50 & 44.50 & 43.59 & 21.38 \\
4zd           &  67 & 0.000 & 1.000 & 0.466 & 41.50 & 43.38 & 42.90 & 21.46 \\
4zf           &  69 & 0.000 & 1.000 & 0.820 & 43.38 & 44.55 & 43.59 & 21.19 \\
\cutinhead{24 $\mu$m-undetected}
1+3$^{\rm a}$ &  58 & 0.000 & 1.000 & 0.767 & 41.50 & 44.50 & 43.19 & 22.44 \\
1$^{\rm b}$   &  44 & 0.000 & 1.000 & 0.780 & 41.50 & 44.50 & 43.33 & 22.52 \\
3$^{\rm c}$   &  14 & 0.000 & 1.000 & 0.684 & 41.50 & 44.50 & 42.96 & 21.18
\enddata
\tablecomments{\small Column (1): sample ID, which matches references to
  specific samples throughout the paper. Column (2): number of
  galaxies in each sample. Columns (3)-(5): minimum, maximum, and
  median redshifts. Columns (6)-(8): logarithm of the minimum,
  maximum, and median 2--10~keV X-ray luminosities, in units of
  erg~s$^{-1}$. Column (9): logarithm of the median column density, in
  units of~cm$^{-2}$.}
\tablenotetext{a}{20 $\le\log(N_{\mathrm{H}}/$cm$^{-2})<$ 24}
\tablenotetext{b}{22 $<\log(N_{\mathrm{H}}/$cm$^{-2})<$ 24}
\tablenotetext{c}{20 $\le\log(N_{\mathrm{H}}/$cm$^{-2})<$ 22}
\end{deluxetable}

\clearpage

\begin{deluxetable}{lccc}
\setlength{\tabcolsep}{0.05in}
\tablewidth{0pc}
\tablecolumns{4}
\tablecaption{Stacking results\label{table:results}}
\tablehead{
  \colhead{ID}  & \colhead{Integrated flux density} & \colhead{S/N} & \colhead{SFR} \\
  \colhead{}    & \colhead{($\mu$Jy)}       & \colhead{}    & \colhead{($\Msun$ yr$^{-1}$)} \\
  \colhead{(1)} & \colhead{(2)}             & \colhead{(3)} & \colhead{(4)}
}
\startdata
\cutinhead{24 $\mu$m-detected, $\log(N_{\mathrm{H}}/$cm$^{-2})>$ 22}
2z  &  15.3 $\pm$  3.0 &  7.6 & 21 $\pm$  4 \\
2za &  21.6 $\pm$  8.9 &  3.2 & 10 $\pm$  4 \\
2zb &  10.6 $\pm$  4.5 &  3.7 & 14 $\pm$  6 \\
2zc &  12.2 $\pm$  4.1 &  4.7 & 24 $\pm$  8 \\
2zd &  21.0 $\pm$  8.9 &  3.1 & 10 $\pm$  5 \\
2ze &  10.2 $\pm$  3.9 &  4.4 & 15 $\pm$  6 \\
2zf &  11.5 $\pm$  4.2 &  4.3 & 20 $\pm$  7 \\
\cutinhead{24 $\mu$m-detected, $\log(N_{\mathrm{H}}/$cm$^{-2})<$ 22}
4z  &  14.1 $\pm$  3.8 &  5.4 & 14 $\pm$  4 \\
4za &  10.2 $\pm$  4.3 &  3.7 &  4.1 $\pm$  1.7 \\
4zc &  18.2 $\pm$  5.7 &  4.5 & 29 $\pm$  9 \\
4zd &   8.1 $\pm$  4.1 &  3.3 &  3.3 $\pm$  1.7 \\
4zf &  19.7 $\pm$  5.1 &  5.5 & 29 $\pm$  8 \\
\cutinhead{24 $\mu$m-undetected}
1+3 & -34.2 $\pm$ 19.9 & \ldots & \ldots \\
1   & -20.0 $\pm$ 10.9 & \ldots & \ldots \\
3   & -39.4 $\pm$ 18.8 & \ldots & \ldots

\enddata
\tablecomments{IDs correspond to the samples listed in Table~\ref{table:samples}.}
\end{deluxetable}

\clearpage

\clearpage

\clearpage

\clearpage

\clearpage

\clearpage

\clearpage

\clearpage

\clearpage

\clearpage

\clearpage

\clearpage

\end{document}